\documentclass[12pt,halfline,a4paper]{ouparticle}
\usepackage{graphicx}
\usepackage{caption}

\usepackage{float}
\usepackage{times}
\usepackage{booktabs}

\usepackage[utf8]{inputenc}
\usepackage[T1]{fontenc}
\usepackage{array}

\newcommand{\PreserveBackslash}[1]{\let\temp=\\#1\let\\=\temp}

\newcolumntype{C}[1]{>{\PreserveBackslash\centering}p{#1}}
\newcolumntype{R}[1]{>{\PreserveBackslash\raggedleft}p{#1}}
\newcolumntype{L}[1]{>{\PreserveBackslash\raggedright}p{#1}}

\begin{document}

\title{Scientific X-ray}

\author{%
\name{Qi Li\footnotemark[2]}
\address{Department of Electronic Engineering, Shanghai Jiao Tong University, Shanghai 200240, China}
\email{liqilcn@sjtu.edu.cn}
\and
\name{Xinbing Wang\footnotemark[2]\ \ \footnotemark[1]}
\address{Department of Computer Science and Engineering, Shanghai Jiao Tong University, Shanghai 200240, China}
\email{xwang8@sjtu.edu.cn}
\and
\name{Luoyi Fu\footnotemark[2]}
\address{Department of Computer Science and Engineering, Shanghai Jiao Tong University, Shanghai 200240, China}
\email{yiluofu@sjtu.edu.cn}
\and
\name{Chenghu Zhou\footnotemark[1]}
\address{Institute of Geographical Sciences and Natural Resources Research, Chinese Academy of Sciences, Beijing 100101, China}
\email{zhouch@lreis.ac.cn}
\newpage
}
\renewcommand{\thefootnote}{\fnsymbol{footnote}} %将脚注符号设置为fnsymbol类型，即特殊符号表示
\footnotetext[2]{These authors contributed equally to this work.} %对应脚注[1]
\footnotetext[1]{Corresponding authors.} %对应脚注[2]

\abstract{The rapid development of modern science has spawned rich scientific topics to research and endless production of literature in them. Just like the widely used X-ray imaging technology, can we intuitively reveal the evolution laws of science from the massive correlation between knowledge? To answer this question, we collect 71 431 seminal articles of topics that cover 16 disciplines and their citation data and propose the corresponding Scientific X-ray framework. We extracts the `idea tree' of each topic by retaining the most representative citation to restore the structure of the development of 71 431 topic citation networks from scratch. At timestamp $t$, we define the Knowledge Entropy ($KE^t(v)$) metric of node $v$, and the contribution of high knowledge entropy nodes to increase the depth of the idea tree (${\Delta D}^t(v)$) is regarded as the basis of topic development. We find two interesting phenomena by Scientific X-ray: (1) Even though the scale of topics may increase unlimitedly, there is an insurmountable upper bound of topic development: the depth of the idea tree does not exceed six-hop. (2) It is difficult for a single article to contribute more than three-hop to the depth of its topic, to this end, the continuing increase of the depth needs to be motivated by the influence relay of multiple high knowledge entropy nodes. We derive a unified quantitative relationship between ${\Delta D}^t(v)$ and $KE^t(v)$: ${\Delta D}^t(v) \approx \log \frac{KE^{t}(v)}{\left(t-t_{0}\right)^{1.914}}$. By transforming the value conditions of the formula and generalize it to the entire topic, we can effectively portray the different patterns of a single topic of depth development and quantify the development potential of topics. Scientific X-ray can be used with few expertise, which can provide important references for grasping research trends, helping policy making and even promoting social development.
}

\date{}


\maketitle

\section*{INTRODUCTION}
\noindent The explosion of the scientific enterprise has boosted the increase of high-quality scientific publication data\cite{mcnutt2014measure, weis2021learning, mukherjee2017nearly, galiani2017life}, significantly attracting interest in utilizing data-driven methods to understand the process of scientific evolution\cite{clauset2017data}. Futurist Raymond Kurzweil, in his futuristic book \textit{The Singularity Is Near}, predicts that human technological development will reach its limit in 2045, i.e. the arrival of the `technological singularity'\cite{kurzweil2005singularity}. At this point, the intelligence of machines has far surpassed humans, causing it difficult for humans to understand the technology created by machines and ultimately making it impossible for humans to push civilization forward. In the current semiconductor industry, as we approach the atomic scale, chip manufacturing has become highly complex, and the increase in revenue becomes less significant\cite{waldrop2016chips}. At present, Moore's Law slowdown has become the industry consensus\cite{itrs2015, aimone2019neural, esmaeilzadeh2011dark}, and soon, the chip miniaturization process will most likely end at 5 nm\cite{leiserson2020there}. Singularity Theory and the slowdown of Moore's Law inspires us to think: Is there a similar life cycle of prosperity and extinction\cite{mane2004mapping, borner2004simultaneous, ke2015defining} in the current scientific evolution?\\

\noindent By observing and counting the citation trends of 68 675 high-impact publications (citation $\geq$ 1000, covering 16 disciplines, published from 1800 to 2019), we find that the impact of publications is dominated by two underlying processes: birth and aging (Fig. 1). The phenomenon of `aging' in high-impact articles indicates that although the scale of the scientific topics they lead are relatively large, the topics will not continue to develop.\\

\begin{figure}[H]
	\centering
	\includegraphics[width=16cm]{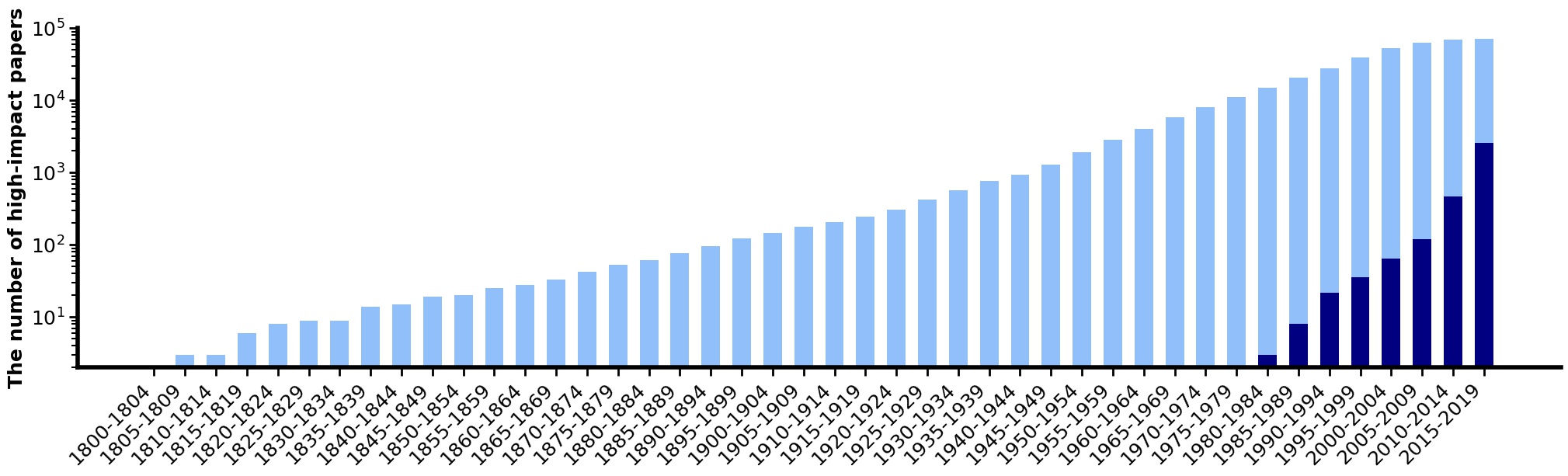}
	\captionsetup{labelfont=bf}
	\caption{Statistics on the moments of `birth' and `aging' of 68 675 high-impact papers. The height of the bar indicates the total number of high-impact articles that have been `born' until certain moment; The dark blue part indicates the total number of high-impact articles that have `aged' until certain moment. The moment when an article `ages' is determined by the citation trend. When a high-impact article has no more than 10 new citations since a certain year, the article is considered to be `aging' from that year; The light blue part of the bar represents the total number of articles that are still active up to the current moment. This figure shows that as high-impact articles continue to appear, some of them will gradually lose attention.}
	\label{fig_1}
\end{figure}

\noindent However, current citation-based indicators\cite{garfield1972citation, garfield2006citation, hirsch2005index, van2014top} only provide a suboptimal reference for the impact evolution of a single academic entity. It is hard to transfer directly from individual academic entities to the scientific topic formed by the interaction of these entities to explore the evolution of the topic. The existence of such complicated correlations leads to the more complex and varied evolution of topics, which makes it extremely difficult to identify specific patterns. Just like X-ray imaging technology, intuitively identifying internal evolution pattern of scientific topics can provide constructive guidance to the decision-making process of resource allocation\cite{weis2021learning, borner2004simultaneous}, researcher hiring\cite{ke2015defining, penner2013predictability}, and direction selection\cite{clauset2017data, mane2004mapping}. For the government, potential scientific topics should be funded to allocate trillions of research funding taxed from the public efficiently\cite{mcnutt2014measure}. For the research institution, the potential value of the hired candidate's research topics should be a key consideration to maximize the recruitment benefit. For a researcher, promising scientific topics should be explored as a priority to publish more impactful work. The depth of understanding of the evolution of scientific topics arguably determines the upper limit of mankind's ability to promote scientific development to achieve more advanced results\cite{ma2018scientific, wang2013quantifying}. However, due to lack of intuitive and in-depth perception of scientific topics, little is known about the evolution laws of them. Here we propose Scientific X-ray framework to provide insight into the evolution of scientific topics.\\
\section*{RESULTS}
\subsection*{The framework of Scientific X-ray}
\subsubsection*{Definition of scientific topic and primary parse}
\noindent To explore the evolution of scientific topics from `birth' to `aging' in depth, we need first to define the scientific topic and seek a suitable way to describe it. In a specific scientific topic, papers usually have similar research interests and are linked by complex citation relationships. Moreover, the birth and summary of the ideas occur among the articles\cite{jia2017quantifying} and diffuses through citations, making the topic developed. Therefore, the scientific topic can be defined and described as a citation network, where nodes represent papers, edges symbolize citations, and the evolution of the network is regarded as the topic's development. High impact papers that break significant ground or create new directions always attract numerous articles with similar research interests. Consequently, the citation network pioneered by a non-trivial influence article and consisting of the leading paper, the articles citing the leading paper and all citations among them can be considered as a particular scientific topic. We collect and integrate academic data from bibliographic databases, including but not limit to IEEE, ACM, arXiv, Elsevier, and Spring. We then select all 71 431 high-impact publications with more than 1000 citations as pioneering works and construct topic citation networks for them. All leading articles were published between 1800 and 2021, and their research interests cover 294 fields in 16 disciplines: History, Computer science, Environmental science, Geology, Psychology, Mathematics, Physics, Materials science, Philosophy, Biology, Medicine, Sociology, Art, Economics, Chemistry, and Political science. To intuitively perceive spatial structure for scientific topics, we visualize all 71 431 topic citation networks to construct the spatial reviews of them, which simulates a gravity system to spatialize a network by structurally determined attraction-repulsive force\cite{jacomy2014forceatlas2}, and we call them galaxy maps. Galaxy map helps us understand the current characteristics of a scientific topic with the power of human vision and spatial cognition by showing the structure of the topic citation network\cite{mane2004mapping}. In Fig. 2(a-d), the two topics' leading work are `The End of History and the Last Man', `A New Statistical Method for Haplotype Reconstruction from Population Data'. The first article, a book of political philosophy published in 1992, presents a argument of `End of History'; the second article, a biological article published in 2001, presents a new statistical method for haplotype reconstruction. By observing the galaxy map we find that the degree of topic development is reflected in the complexity of the structure of topic citation network, and the increase in complexity is driven by high impact nodes. Although the monograph `The End of History and the Last Man' has accumulated more than 6 000 citations, the structure of the topic network has not evolved to become more complex, suggesting that the topic has not moved forward (Fig. 2(b)). In contrast, in the\\

\begin{figure}[H]
	\centering
	\includegraphics[width=16cm]{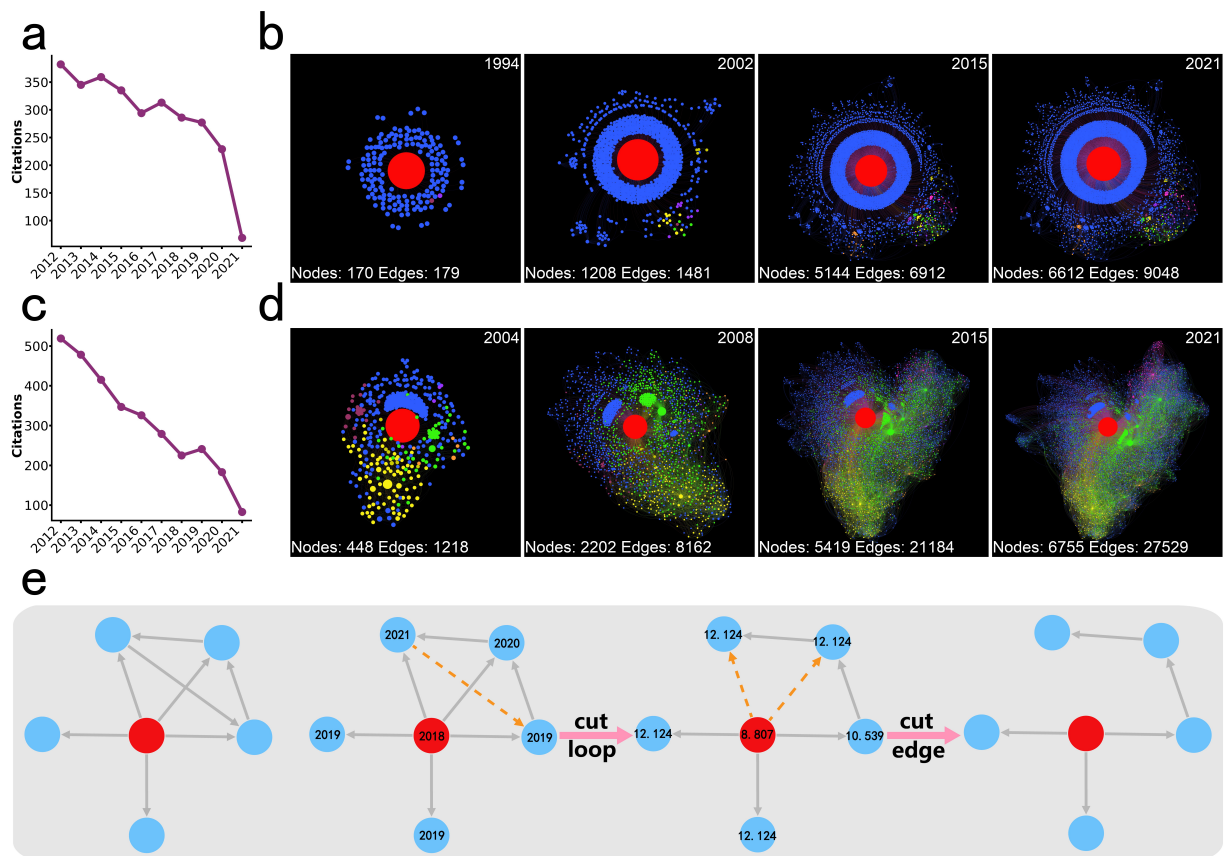}
	\captionsetup{labelfont=bf}
	\caption{(a,c) Citation trends of `The End of History and the Last Man' and `A New Statistical Method for Haplotype Reconstruction from Population Data' in the last decade. New citations of the two topics is declining year by year. (b,d) The evolution of the galaxy map led by `The End of History and the Last Man' and `A New Statistical Method for Haplotype Reconstruction from Population Data'. The red node in the galaxy map is the leading article of the topic. Except for the blue nodes, other nodes with the same color represent that they belong to a larger community, and the blue node does not belong to these communities. The size of the node is positively related to its citation within the topic. The difference in the development of the two topics is reflected in the complexity of the galaxy maps' structure. (e) The diagram of idea tree extraction. The direction of the arrow represents the direction of the flow of ideas, which is opposite to the direction of citations. The numbers on the nodes respectively represent the year of publication of the corresponding articles and their reduction index to the entire network. The algorithm first removes the directed loops in the data to ensure that the paper published priorly is always cited by the papers published later. Then the algorithm cuts off the redundant edges based on the similarity (determined by the reduction index to the entire network between two nodes) between the nodes to keep the most important citation for each node. (See the detailed description of the algorithm in the Methods section and DERIVATION DETAILS section of Supplementary Materials.)}
	\label{fig_2}
\end{figure}

 \noindent  topic led by the article `A New Statistical Method for Haplotype Reconstruction from Population Data', several high-impact child nodes are born during its evolution (Fig. 2(d)). After their children nodes inheriting and developing leading article's idea, they will attract new attention for the development of the topic to form a more complex community structure, which  indicates that the topic has achieved substantial development. Similarly, these two topics are also `aging' in the later stages of their development. In terms of citation trends, the number of new citations for both topics is decreasing year by year (Fig. 2(a,c)), as the topics are reaching the limits of their development due to the difficulty of breeding new high-impact children nodes to break the silence. \\

\subsubsection*{Extraction of scientific topic idea tree}

\noindent Although galaxy maps can transform abstract data into intuitive structures, they contain too much redundant information that makes it difficult to clearly reveal the evolution laws of scientific topics. The development of topics is driven by the inheritance of existing knowledge and the creation of new knowledge through newly published work. In contrast, the structural redundancies in the network generated by non-critical citations have a minimal effect on the development of topics. Therefore, repetitive, invalid inheritance relationships need to be removed to clearly and accurately reproduce the pattern of topic evolution, which can be achieved by assessing the similarity between papers. Ideally, we assume that any child article in the topic except the seminal work is inspired by one of the most essential citation so that we can get an `idea tree' that describes the backbone of knowledge flow within the topic and reveals the inheritance of ideas between different articles (Fig. 2(e))\cite{fu2021can}. (See the detailed description of the algorithm in the Methods section and DERIVATION DETAILS section of Supplementary Materials.) The idea tree is similar to a tree in nature in different types of work play different roles: the seminal work is like the root of the tree that provides the impetus for the development of the topic with their unparalleled innovation; the non-trivial work with different values of knowledge is like the trunks of the tree that drive the growth of their branches; the ordinary articles in the topic are more like the leaves that nourish the impact of the trunk work. Within the topic, when existing knowledge flows through the nodes, new knowledge is generated after cognition, fusion, and creation by the human brain and then inspires later generations. These newly added nodes tend to be more similar to the creators of new knowledge than the propagators of the original knowledge. The difference in the quality of the new knowledge leads to a difference in the number of new child nodes attracted, which is expressed macroscopically as a difference in the structure of the idea tree. In this way, we can characterize different evolution patterns through different idea tree structures. Based on the topic idea tree we got, we can reproduce the evolutionary pattern of the topic and thus explore the inner causes of the limits by tracing the main paths of inheritance of ideas in the topic.\\

\subsubsection*{Measurement of knowledge quality and definition of visible depth}

\noindent The formation of idea tree structure is driven by nodes with different knowledge qualities. The emergence of new, innovative, and inspiring knowledge can breathe new life into the topic with the attraction of additional research, which is the critical internal factor that stimulates the evolution of inheritance relationships. Therefore, the effective measurement of knowledge is a fundamental question in revealing the evolution of scientific topics. Although knowledge exists objectively, it is as invisible as air, making it difficult to define and lacks a unified quantification. However, the impact of knowledge can be felt all the time: the more knowledge a person has, the more complex work he or she can do, and the more knowledge a machine can learn, the more intelligent it can get. For publications, it is almost impossible to go directly, effectively, and uniformly evaluate the quality of knowledge without any experience. Nevertheless, the idea tree preserves the most useful information within the topic through a process of de-redundancy, and differences in the complexity of the subtree structure led by a node characterize the difference in the quality of knowledge at that node. Although knowledge is hard to capture, based on the idea tree whose structure is shaped by the flow of knowledge through the topic network, we can use structural information to measure and analyze the quality of knowledge. Structural information has proven to be very useful in analyzing and measuring the complexity of network structures, which is widely used in graph data analysis of computer science\cite{liu2021bridging, chen2019fast, minello2019neumann}, biology\cite{li2018decoding,bae2008bioinformatic,garcia2015rna} and physics\cite{de2016spectral, girolami2017quantifying, arola2018synchronization}. Here we have been inspired by previous work\cite{li2016structural} to design a knowledge entropy to measure the quality of knowledge in the idea tree nodes. Based on this, we can describe at a fine-grained level the contribution of different qualities of knowledge to the development of the topic. At timestamp $t$, the topic citation network is $G^t=\left(V^t,E^t\right)$. The idea tree extracted from the network is $IdeaTree\left(G^t\right)$. For any paper $v$ belong to $IdeaTree\left(G^t\right)$ except the seminal work, we define its knowledge entropy ${KE}^t\left(v\right)$ as:
\begin{equation}
  {KE}^t\left(v\right)=\left|H^t\left(v\right)-\sum_{v_i\in C^t\left(v\right)}{H^t\left(v_i\right)}+\sum_{v_i,v_j\in C^t\left(v\right),i\neq j}{I^t\left(v_i,v_j\right)}\right|  
\end{equation}
\noindent where $H^t\left(v\right)$ represents the subtree entropy of the subtree led by node $v$ at $t$, $C^t\left(v\right)$ represents the children of $v$ in $IdeaTree\left(G^t\right)$ at $t$, and $I^t\left(v_i,v_j\right)$ represents the mutual subtree entropy of the subtree led by $v_i$ and $v_j$ at $t$. (See the detailed derivation of the formula in the Methods section and DERIVATION DETAILS section of Supplementary Materials.)\\

\noindent Knowledge changes as it is passed on. For any child node in the idea tree, each time an idea is inherited along the connection path from the seminal work to this paper, a difference in knowledge is accumulated. In terms of the whole topic, when we use pioneering work as a reference, the depth of the idea tree represents the difference between the new knowledge and the original knowledge, which characterizes the degree of the development of the topic. Besides, knowledge entropy helps us to identify it. The appearance of a high knowledge entropy node at a specific level of the idea tree symbolizes a qualitative shift in the accumulation of knowledge variation, resulting in the emergence of knowledge of extraordinary value within the topic, thus marking a substantial development of the scientific topic. Based on this, we call the layer in the idea tree that contains at least one node with knowledge entropy greater than $M$ visible layer. With the development of the topic, the deeper layers will gradually become visible. Therefore, based on the idea tree and knowledge entropy, at timestamp $t$, we define the number of visible layers as the visible depth $\left(VD^t\right)$ of the topic to characterize the degree of development of the topic.\\

\subsection*{Measuring the development potential of an article/topic}
\noindent Can we quantify the driving effect of high knowledge entropy nodes on visible depth to predict the evolution of visible depth? We can judge and predict the vitality and development potential of the topic based on the structure of the idea tree. We find that nodes with small knowledge entropy are hard to stimulate depth. When the knowledge entropy does not increase to inhibit the development of the topic (see the detailed introduction of the phenomenon in the The evolution patterns of scientific topics section), the greater the knowledge entropy is, the stronger the node's ability to stimulate depth. As mentioned above, the ability of a node to stimulate depth is positively related to its knowledge entropy. However, there are still cases where some nodes with high knowledge entropy fail to stimulate depth. We find that these nodes tend to have high knowledge entropy in the early stage but fail to drive topic depth increase even though the duration is significantly longer than the time interval of visible depth increase. This suggests that the increase in visible depth tends to occur within an effective duration. If a high knowledge node fails to increase the depth for a more extended period, it will become less attractive for additional research. It will gradually lose the opportunity to drive the topic forward. The increase in visible depth results from the combined effect of increased knowledge entropy and time, and we believe that time begins to decay the node's ability to stimulate depth from the influence of the paper beginning to emerge. In contrast, when the effect of increasing knowledge entropy is stronger than the attenuation effect of time, it will increase visible depth. However, the decay effect of time is always present, similar to how the radioactivity of isotope attenuates over time in nature. Since the knowledge entropy of nodes will not keep increasing, in the long run, the driving effect of high knowledge entropy nodes will end in the future, and then the development of the topic will come to a standstill. We derive the Idea Limit Formula for quantitatively describing the effect of nodes on driving topic development based on conjecture, fitting, and verification, which can quantify the development potential of the article within the topic. (See the fitting detail of the formula in the Methods section.) Assume that the minimum knowledge entropy of the nodes makes a certain level of the idea tree visible $KE_{Threshold}=M$. Node $v$ becomes visible at $t_0$, i.e. $KE^t_0\left(v\right)\geq M$. The depth ${\Delta D}^t(v)$ that the node $v$ can stimulate for the topic after time $t$ satisfies:
\begin{equation}
{\Delta D}^t(v) \approx \log \frac{KE^{t}(v)}{\left(t-t_{0}\right)^{1.914}}
\end{equation}
\noindent Furthermore, based on the idea limit formula of a single high knowledge entropy node, we can predict the increase in the visible depth of the topic ($\Delta D_{Topic}^{t}$) in the future after the moment $t$. Let $S^t$ be the set of all nodes in the idea tree that meet $KE^t\left(v\right)\geq M$, the visible depth of the current idea tree is $VD^t$, for any $v_{i} \in S^{t}$, its visible depth in the idea tree is $V D_{v_{i}}^{t}$, according to the idea limit formula, the visible depth that can be stimulated by node $v_i$ for the topic after the moment $t$ is $\Delta D^{t}\left(v_{i}\right)$. Therefore, for the whole topic we use $\Delta D_{Topic}^{t}$ to quantify its development potential:
\begin{equation}
\Delta D_{Topic }^{t}=\max _{v_{i} \in S^{t}}\left\{\Delta D^{t}\left(v_{i}\right)-\left(V D^{t}-V D_{v_{i}}^{t}\right)\right\}
\end{equation}

\noindent \subsection*{The six-hop upper bound of any scientific topic development}
\noindent Through the statistics of maximum visible depth for 3 953 topics selected randomly from 71 431 scientific topics in the entire field (Fig. 3(a)), we find that even though the number of publications of topics may increase unlimitedly, there is an insurmountable upper bound of topic development: The visible depth of 99.9\% of scientific topics is difficult to exceed six-hop, which coincides with the `Six Degrees of Separation' theory\cite{milgram1967small} of psychology and the `small-world' theory\cite{watts1998collective} of network science. We find that the number of topics with a limit depth of zero is the largest (30.89\%, Fig. 3(a)), suggesting that it is difficult for topics to make the breakthrough from zero to one. The decrease in the number of topics with depth increasing indicates that the later the topic is developed, the more difficult it is for valuable knowledge to emerge.\\

\subsection*{The evolution patterns of scientific topics}
\noindent We utilize Scientific X-ray to explore the evolution patterns of topics. Among all topics, six representative topics covering the fields of geographic information system, ecology and climate change, computer vision, natural language processing, deep learning and geology have been selected to reveal depth evolution dynamics. We find that topic evolution follows six fixed patterns through our through our induction and summary, which are widespread in all disciplines. (More example, see EXAMPLES OF SIX EVOLUTION PATTERNS section in Supplementary Materials.)\\

\begin{figure}[H]
	\centering
	\includegraphics[width=16cm]{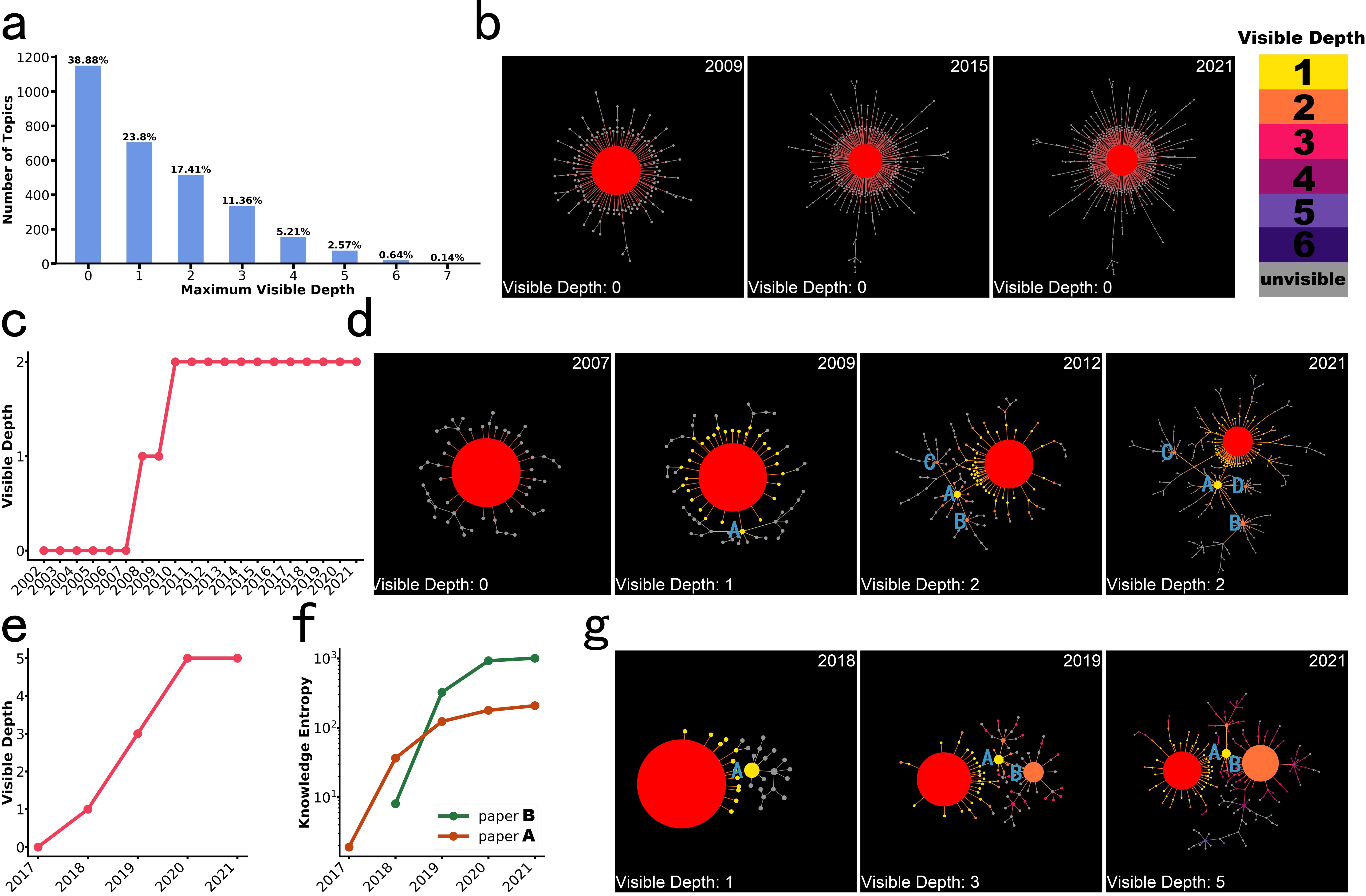}
	\captionsetup{labelfont=bf}
	\caption{(a) Maximum visible depth distribution of scientific topics. 99.9\% of scientific topics are difficult to exceed six-hop. (b,d,g) The evolution of the idea trees led by `Geographic Information Systems and Science', `Range Shifts and Adaptive Responses to Quaternary Climate Change' and `Multi-view 3D Object Detection Network for Autonomous Driving'. The idea tree is visualized by the ForceAtlas2 algorithm\cite{jacomy2014forceatlas2}. The color of the node represents its visible depth, and node size is positively related to its knowledge entropy. (c,e) The evolution of the visible depth of the topics led by `Range Shifts and Adaptive Responses to Quaternary Climate Change' and `Multi-view 3D Object Detection Network for Autonomous Driving'. (f) The evolution of the knowledge entropy of the corresponding nodes within the topics led by `Multi-view 3D Object Detection Network for Autonomous Driving'.}
	\label{fig_3}
\end{figure}

\noindent \textbf{$\bullet$\quad Pattern 1:} The visible depth of the topic pioneered by summative work is hard to exceed one. The leading work of the topic shown in Fig. 3(b) is `Geographic Information Systems and Science', which is a textbook of Geographic Information System (GIS) published in 2001. The scale of such type of topics grows over time, but their visible depth barely increases. Pioneering work tends to summarize existing knowledge instead of proposing new methods or theories, so it is not very inspiring for child nodes and cannot provide new research ideas. We find that almost all idea trees led by summative work such as textbooks, surveys and software toolkits show a similar structure.\\
\noindent \textbf{$\bullet$\quad Pattern 2:} The increase in visible depth needs to be driven by non-trivial child nodes. The leading work of the topic shown in Fig. 3(c,d) is `Range Shifts and Adaptive Responses to Quaternary Climate Change'. The lead article presents that the core of the biotic response to climate change lies in the interaction of adaptation and migration, and discusses how rapid climate change influences regional ecosystem and species diversity. The child article A: `Constraint to Adaptive Evolution in Response to Global Warming' is inspired by the leading article and further proposes that under the conditions of global climate warming, the genetic architecture of three populations of a native North American prairie plant is much slower than climate change. Paper A has attracted a lot of outside attention, making subtree led by it flourish and giving birth to three new high knowledge entropy articles including:  `Predicting the impacts of climate change on the distribution of species: Are bioclimate envelope models useful?' (paper B), `Evolutionary Responses to Climate Change' (paper C) and `Adaptation, migration or extirpation: climate change outcomes for tree populations' (paper D). All these papers continue to discuss the impact of climate change on biological evolution and attract a lot of attention, which increasing the visible depth of the topic to two (Fig. 3(c)). The article `Constraint to Adaptive Evolution in Response to Global Warming' inherits the knowledge of the leading article and stimulates the generation of new valuable knowledge, thereby promoting the development of the topic. \\

\noindent \textbf{$\bullet$\quad Pattern 3: }The continuous increase of the visible depth of the topic needs to be stimulated by the influence relay of multiple high knowledge entropy nodes. The leading work of the topic shown in Fig. 3(e-g) is `Multi-view 3D Object Detection Network for Autonomous Driving'. The lead article creatively fuses visual information with radar point cloud data and applies it to 3D object detection in autonomous driving to enhance the model. Child article a: `3D fully convolutional network for vehicle detection in point cloud' is inspired by the lead article, which introduces a 2D fully convolutional network into a 3D point cloud, thus enabling 3D object detection. Its influence started to show in 2018, but the growth of its knowledge entropy slowed down significantly after 2019. And at this time, the child node B directly inspired by paper A: `Frustum PointNets for 3D Object Detection from RGB-D Data' assists the object detection in 3D point cloud data with 2D image detection information, which enables the model to achieve high accuracy 3D object detection for indoor and outdoor scenes. This allows its knowledge entropy to surpass that of article A and takes over the task of motivating the visible depth increase (Fig. 3(f)). Inspired by article B, the topic can push forward and continue to attract outside research interest. This leads to the spawning of several new high knowledge entropy child nodes under the subtree led by article B and makes the visible depth of the topic continue to increase to five (Fig. 3(e)).\\

\noindent \textbf{$\bullet$\quad Pattern 4: }The presence of overpowered child nodes can ruin the increase in the visible depth of the topic. The leading work of the topic shown in Fig. 4(a-c) is `Deep contextualized word representations'. The leading article proposed ELMo, a word vector model that can learn contextual features. The model achieves significant performance improvements on multiple tasks and datasets, and the work was awarded the Outstanding Paper Award at NAACL 2018. However, the leading article was overshadowed by the emergence of BERT (article A), which changed the game by moving word representations toward larger models. The knowledge entropy of the child article A: `BERT: Pre-training of Deep Bidirectional Transformers for Language Understanding' rapidly increased by over ${10}^3$ orders of magnitude, thus approaching\\

\begin{figure}[H]
	\centering
	\includegraphics[width=16cm]{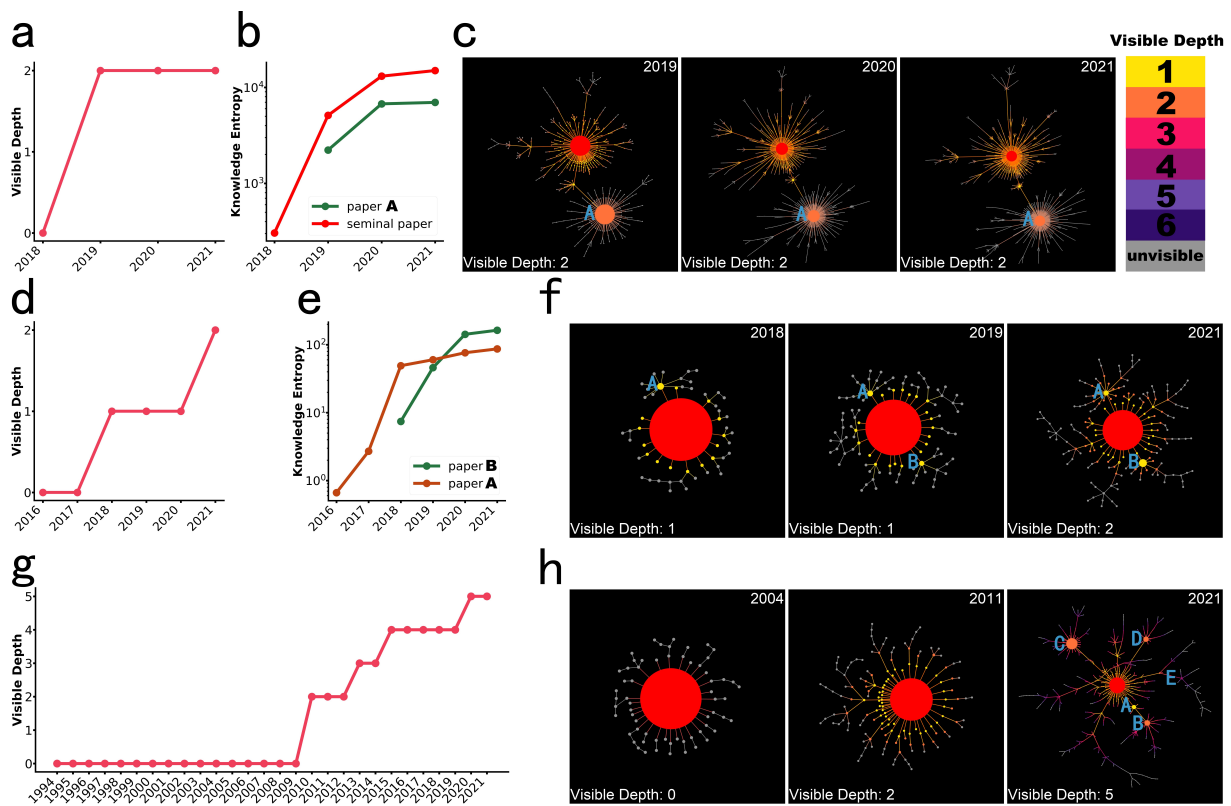}
	\captionsetup{labelfont=bf}
	\caption{(a,d,g) The evolution of the visible depth of the topics led by `Deep contextualized word representations', `DoReFa-Net: Training Low Bitwidth Convolutional Neural Networks with Low Bitwidth Gradients' and `Evolution of the Altaid tectonic collage and Palaeozoic crustal growth in Eurasia'. (b,e) The evolution of the knowledge entropy of the corresponding nodes within the topics led by `Deep contextualized word representations' and `DoReFa-Net: Training Low Bitwidth Convolutional Neural Networks with Low Bitwidth Gradients'. (c,f,h) The evolution of the idea trees led by `Deep contextualized word representations', `DoReFa-Net: Training Low Bitwidth Convolutional Neural Networks with Low Bitwidth Gradients' and `Evolution of the Altaid tectonic collage and Palaeozoic crustal growth in Eurasia'. The idea tree is visualized by the ForceAtlas2 algorithm\cite{jacomy2014forceatlas2}. The color of the node represents its visible depth, and node size is positively related to its knowledge entropy.}
	\label{fig_4}
\end{figure}

\noindent the order of magnitude of the seminal article(Fig. 4(b)). This indicates that he has become a new authority, which makes most of the articles citing it do not cite the lead article of the original topic, and thus the original topic is deprived of new valuable knowledge and eventually stagnates in development(Fig. 4(a)).\\

\noindent \textbf{$\bullet$\quad Pattern 5: }Stronger branches within the topic inhibit the increase in visible depth of weaker branches. 
The leading work of the topic shown in Fig. 4(d-f) is `DoReFa-Net: Training Low Bitwidth Convolutional Neural Networks with Low Bitwidth Gradients'. The lead work proposes a method that can train convolutional neural networks with low bit-width weights and activation values using low bit-width parameter gradients, which makes it possible to accelerate the training of low-bit networks in hardware. In the next layer of leading work, two nodes with high knowledge entropy were born, namely `Quantized Neural Networks: Training Neural Networks with Low Precision Weights and Activations' (Article A) and `ShuffleNet: An Extremely Efficient Convolutional Neural Network for Mobile Devices' (Article B). These two articles formed two factions within the subject. Article A introduced the concept of Quantized Neural Networks (QNN) and showed that higher computational efficiency can be achieved in hardware when compressing the 32-bit full-precision data of a neural network into 6 or 8-bit floating-point numbers. Early in the development of the topic, the knowledge entropy by article A was the first to grow (Fig. 4(e)). However, article B proposes a very different approach to improve the efficiency of neural networks than article A, which achieves model compression and acceleration by designing a more efficient network structure rather than compressing a large trained model, which enables ShuffleNet to be deployed in resource-constrained mobile deployments. It is the extremely high research and application value of article B that makes its knowledge entropy surpass that of article B after 2018 (Fig. 4(e)), and accordingly, the growth rate of knowledge entropy of article A starts to slow down (Fig. 4(e)). In this process, article B attracts outside attention, making the branch led by article a neglected, thus causing its development to stagnate.\\

\noindent \textbf{$\bullet$\quad Pattern 6: }Visible depth near the upper bound of development requires a large number of high knowledge entropy nodes to drive. The leading work of the topic shown in Fig. 4(g,h) is `Evolution of the Altaid tectonic collage and Palaeozoic crustal growth in Eurasia'. The leading article proposes a new tectonic model and shows that Asia grew by 5.3 million square kilometres during the Palaeozoic era. Inspired by the leading article, a large number of high knowledge entropy papers studying the structural evolution of the Central Asian continent have appeared in the topics, such as `Mesozoic tectonic evolution of the Yanshan fold and thrust belt, with emphasis on Hebei and Liaoning provinces, northern China' (paper A), `Accretion leading to collision and the Permian Solonker suture, Inner Mongolia, China: Termination of the central Asian orogenic belt' (paper B), `The Central Asian Orogenic Belt and growth of the continental crust in the Phanerozoic' (paper C), `Paleozoic accretionary and collisional tectonics of the eastern Tianshan (China) : Implications for the continental growth of central Asia' (paper D) and `Structural constraints on the evolution of the Central Asian Orogenic Belt in SW Mongolia' (paper E). The knowledge entropy of all these articles exceeds $10^2$. Because of this, the visible depth of the topic can reach five (Fig. 4(g)), which also reflects the great innovation and inspiration of leading the work from the side.\\

\noindent By transforming the value conditions of the formula (2), we can effectively portray the different\\

\begin{figure}[H]
	\centering
	\includegraphics[width=16cm]{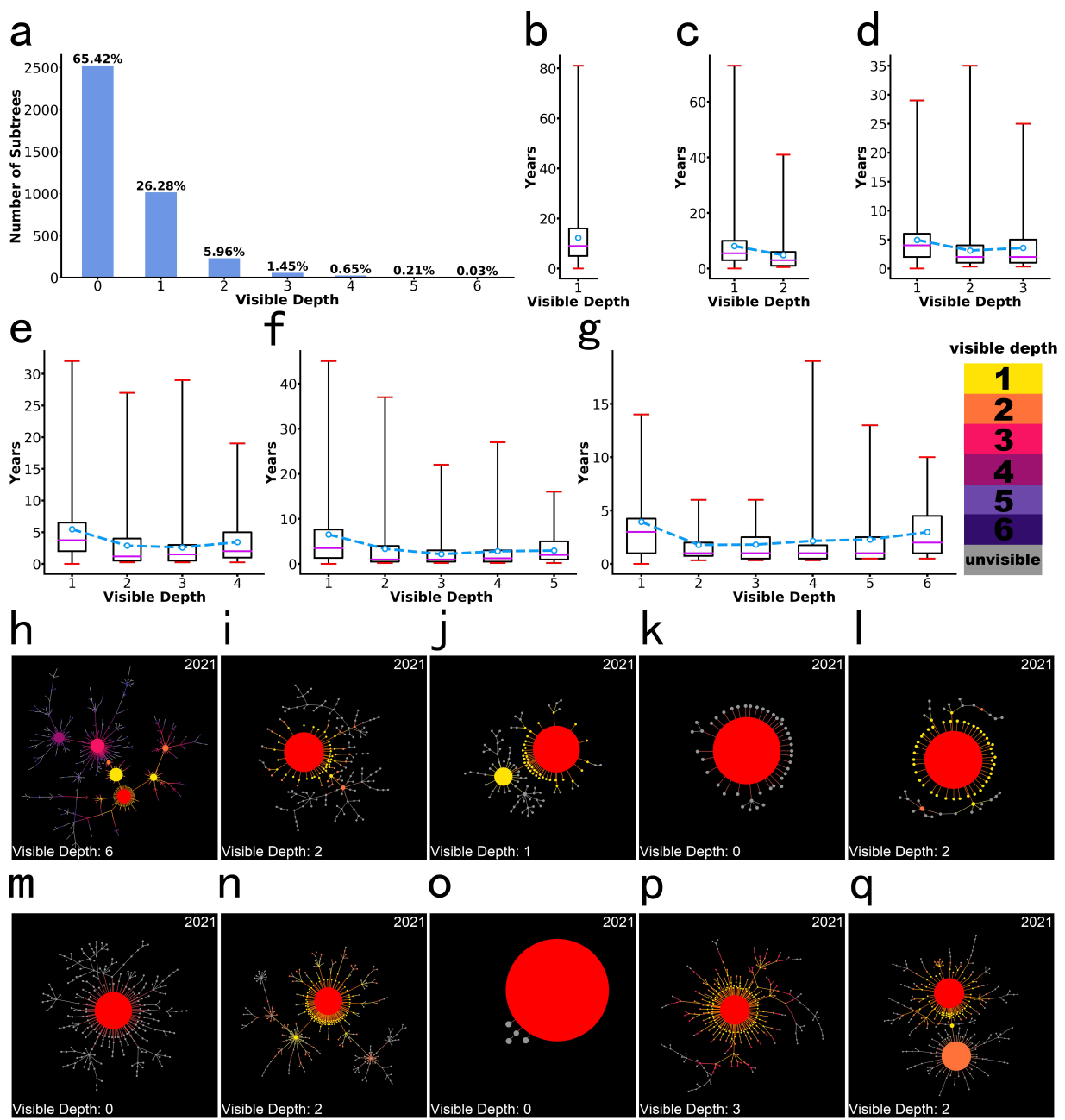}
	\captionsetup{labelfont=bf}
	\caption{(a) The distribution of the visible depth inspired by a single article within topic. The idea limit of any single article is hard to exceed three-hop. (b-g) The time interval distribution of the topic visible depth change with the maximum visible depth of one to six. The blue dotted line represents the trend of average time to germinate a new layer. The topic depth is difficult to achieve a breakthrough from zero to one. Due to the limit of a single node in promoting the development of the topic, all types of topics with a maximum visible depth greater than two show an increase in the average time to germinate a new layer in the late stages of development. (h) The structure of the idea tree led by `A programmable dual-RNA-guided DNA endonuclease in adaptive bacterial immunity' in 2015 (The Champion of science’s 2015 breakthrough). (i-q) The idea trees of runner-up topics of science’s 2015 breakthrough. The structure of the idea tree of the champion CRISPR is much more complicated than other topics.}
	\label{fig_5}
\end{figure}

\noindent  patterns of a single topic in depth evolution: (1) Too small knowledge entropy cannot advance the topic depth to achieve a breakthrough (${\Delta D}^t(v)$ is always less than 1), which can explain the situation that the topic cannot breed new high knowledge entropy nodes and then reaches the development cap. (2) If the nodes with high knowledge entropy do not promote the increase of the visible depth within the effective time, the topic development will stagnate. This can describe the situation where the topic has bred high knowledge entropy nodes, but the depth has not been increased for a long time. It also includes Pattern 4 mentioned above. (3) If the high knowledge entropy nodes appearing in the topic are still in the prime time to stimulate the depth of the topic, i.e. the upper bound of ${\Delta D}^t(v)$ is greater than 1 when specific values are brought into the formula, we consider the topic to have a very high potential for development.\\

\subsection*{The idea limit of any single article within the topic}
\noindent When articles with different knowledge values appears in different positions of the idea tree, it will inspire different subtree structures. Combining subtrees with different shapes can get the skeleton structure of the topic, corresponding to the different evolution patterns of the topic. Therefore, we conjecture that the existence of development upper bound of scientific topics is due to the limited ability of nodes to drive visible depth. To verify this conjecture, we randomly take out subtrees led by high knowledge entropy nodes ($KE \geq 10$) totaling 2 948 that do not contain each other to analyze the driving effect of nodes with different knowledge entropy on topic development. We find that 99.11\% of the articles have difficulty contributing more than three-hop to the visible depth of their topic (Fig. 5(a)), to this end, the continuing increase in the depth of idea tree needs to be motivated by the influence relay of multiple high knowledge entropy nodes , which leads to the emergence of Pattern 3 in the section of The evolution patterns of scientific topics. Moreover, when the knowledge entropy of the article increases to exceed the leading work, due to the destruction effect in Pattern 4, this shifts its effect on the visible depth from facilitation to inhibition, fundamentally leading to the limit of the depth that a single article can drive. The existence of this limit makes it necessary to spend additional time breeding new valuable knowledge to drive the increase of visible depth when the topic reaches the maximum visible depth that a single high-impact article can inspire. We respectively counted the time interval of topic visible depth change for the topics whose maximum visible depths are range from one to six. We find that the visible depth of all types of topics took the longest time to develop from zero to one, verifying that the topic depth is difficult to achieve a breakthrough from zero to one (Fig. 5(b-g)). All types of topics with a maximum visible depth greater than two show an increase in the average time to germinate a new layer in the late stages of development (Fig. 5(d-g)). This is because nodes with extraordinary influence appearing early in the topic reach the limit of their driving effect on the topic development. The topic takes longer to germinate new valuable knowledge.\\

\begin{figure}[H]
	\centering
	\includegraphics[width=16cm]{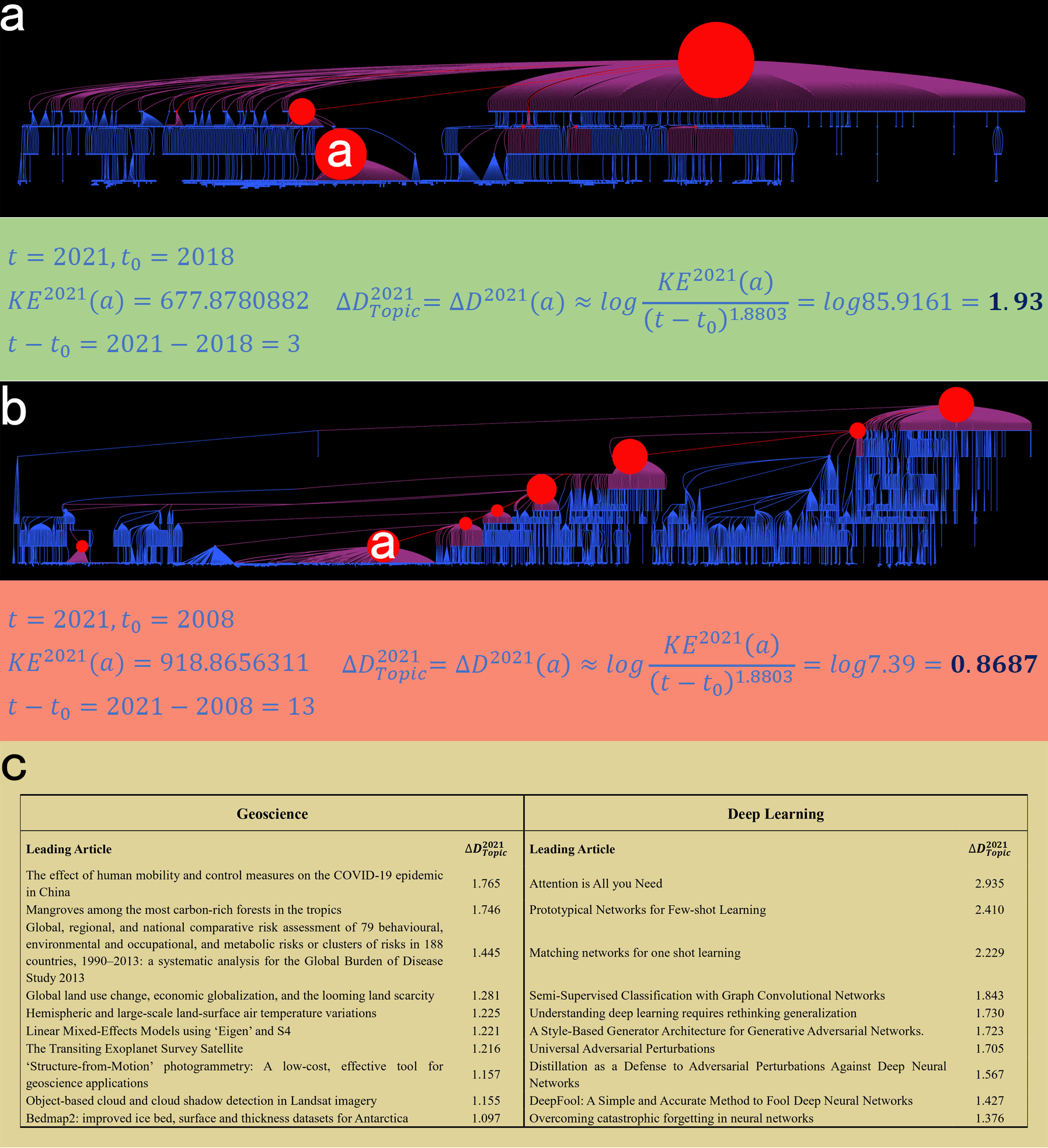}
	\captionsetup{labelfont=bf}
	\caption{(a, b) Use Scientific X-ray to quantify the development potential of the topics led by `Image Super-Resolution Using Deep Convolutional Networks' and `Network information flow'. The idea tree is visualized by the DOT algorithm\cite{gansner1993method}, and node size is positively related to its knowledge entropy. The topic led by `Image Super-Resolution Using Deep Convolutional Networks' has more development potential. (c) Top ten topics in the field of geoscience appearing in the past five years and top ten topics in the field of deep learning appearing in the past ten years according to $\Delta D_{Topic }^{2021}$.}
	\label{fig_4}
\end{figure}
\subsection*{The parse of science’s 2015 breakthrough}
 \noindent We parse the topics covered in the 10 advances in science’s 2015 breakthrough\cite{sci2015breakthrough} to validate the ability of Scientific X-ray in cross-disciplinary impact comparisons. We selected the citation data of these topics in 2021 for parsing. The first place on the list is CRISPR (Clustered Regularly Interspaced Short Palindromic Repeats) gene editing technology. As shown in Fig. 5(h-q), we find that the idea tree structure of CRISPR (Fig. 5(h)) is more fleshed out than the runner-up topics (2-10 of the top 10 advances, Fig. 5(i-q)). CRISPR was also selected as science’s breakthrough in 2012 and 2013 and won the Nobel Prize in Chemistry in 2020, which further highlights its far-reaching impact unlike other topics. We can intuitively perceive the impact of CRISPR based on the parse results of Scientific X-ray without excessive professional knowledge and other additional analysis.\\
 
\subsection*{Measuring the development potential of specific topics}
\noindent Based on the predictive power of Scientific X-ray for topics with development potential, we can identify scientific topics with potential research value, even without any prior expertise. This  has significant implications for the decision-making process of resource allocation, researcher hiring, and direction selection. We selected two scientific topics in computer science, including and deep learning and network information theory. Then we utilize Scientific X-ray to identify whether the scientific topics still have the potential for development (Fig. 6(a,b)). The topics are led by `Image Super-Resolution Using Deep Convolutional Networks' and `The capacity of wireless networks' respectively. These topics have produced a great deal of high-impact work and valuable knowledge throughout their development. In the topic led by `Image Super-Resolution Using Deep Convolutional Networks', After calculating its $\Delta D_{Topic}^{t}$ by formula (3), we finally obtain that the $\Delta D_{Topic}^{2021}=1.918$ (Fig. 6(a)), which is greater than 1. This indicates that the subtree led by it is in a prime period of development and deserves continued research investment, which promises to produce impactful work and drive the depth of the topic. However, in our experience, the topics led by `The capacity of wireless networks' has nearly reached its cap of development. Almost all important issues have been solved in this topics so that a complete knowledge system has been formed. We also obtain that the $\Delta D_{Topic}^{2021}$ is 0.831 (Fig. 6(b)), which indicates that the visible depth of the topic is difficult to increase and coincides with our cognition. Due to the passage of time, the attractiveness of the valuable knowledge in the topic started to decline, leading to a stagnation of the topic development. We also quantify the development potential of scientific topics in different fields including geoscience and artificial intelligence (Fig. 6(c)). (The measurement of topics' development potential in the field of artificial intelligence is divided into deep learning, computer vision, natural language processing and data mining separately. The detailed description of the other three sub-fields can be found in the Supplementary Materials S4.) Because of the different development speeds of the two fields, we set the statistical time in the field of geoscience to recent ten years and the field of artificial intelligence to recent five years. We find that Scientific X-ray can even capture research hotspots in different fields with few expertise, such as `COVID-19 spatio-temporal data mining' in geoscience and `Attention mechanism' and `Few-shot Learning' in the field of deep learning, which can help researchers in different disciplines grasp research trends. (See detailed description of deep learning and geoscience in Supplementary Materials S4.)\\

\section*{DISCUSSION}
\noindent Thanks to the rapid development of current deep learning techniques, we can extract structural and semantic features of articles more accurately, which will improve the models' ability to identify the limits and evolutionary patterns of topics and judge their development potential. However, the improvement in the ability to quantify knowledge relies on a deeper understanding of how knowledge itself is generated and how it interacts with each other. There is an obvious limitation in quantifying knowledge from the structural perspective: we have difficulty in measuring the knowledge of articles with extremely sparse structure. For a valuable article, due to a short publication time or being ignored for a long time for some reason, the quality of knowledge calculated from the structure will deviate from its own. Yet, knowledge structure is the external expression of knowledge under complex interaction mechanisms, and the quality of knowledge contained in an article will not change since its publication and is not related to the complexity of knowledge structure. Therefore, the further improvement of model's capability relies on how to achieve an effective measurement of knowledge quality in the early stage of article publication.\\
\section*{METHODS}

\subsection*{Idea tree extraction algorithm}
\noindent In academic networks, it is rare for two articles to cite each other, because the publication order of the articles is usually sequential. But in actual data, we found very few such data. For this kind of data, we follow the principle that the articles  published later cite the article published prior, which will remove the directed loop in the network. We then extract the idea tree from the topic citation network. The algorithm consists of three steps. Initially, distance between nodes in high-dimensional space are measured via reference relationships. As for a topic citation network, we noticed that the pioneering paper cites none of the papers in the research work since it's the earliest one, which blocks our later calculation for eigenvalues and eigenvectors. Considering this, we involve self-citation or self-loop to the pioneering paper, which allows subsequent eigenvalue decomposition. After the process above, we get the adjacency matrix $W$ with self-loop and the out-degree matrix $D$ with self-loop. Then we get the normalized Laplace matrix with self-loop:\\
\begin{equation}
  L_{normal}=D^{-\frac{1}{2}}(D-W)D^{-\frac{1}{2}}
  \tag{M1}
\end{equation}

\noindent Based on this, we perform spectral decomposition of $L_{normal}$ and get the embedding $eigvector_v$ of any node $v$ in the high-dimensional space. Therefore, for any two paper $v_i$ and $v_j$, we can the calculate the distance of the papers in the high-dimensional space.\\
\begin{equation}
  d_{ij}=\left\Vert eigvector_{v_i}-eigvector_{v_j}\right\Vert _{2}
  \tag{M2}
\end{equation}
\noindent For nodepair $(v_i,v_j)$, to measure the difference between  $v_i$ and $v_j$, we define the Reduction Index of nodepair as the sum of the weighted dijkstra path from $v_i$ to all $v_{j_{k}}$ in $v_j$'s reference list.\\
\begin{equation}
  Reduction Index_{v_i,v_j}=\underset{v_{j_{k}}}{\sum}dijkstra_{v_i,v_{j_{k}}}
  \tag{M3}
\end{equation}
\noindent For node $v$, its Reduction Index to the Entire Network $G(V,E)$ is defined as the sum of its Reduction Index to all other nodes in the network.\\

\begin{equation}
  Reduction Index_{v,G}=\underset{u \in V/{v}}{\sum}Reduction Index_{v,u}
  \tag{M4}
\end{equation}

\noindent Reduction Index to the Entire Network helps us judge the importance of citations. The greater of the difference in the reduction index of two nodes to the network is, the less important the reference relationship between them is. Therefore, we cut the unimportant citations with difference in Reduction Index to the network while maintaining the connectivity in Directed Graph conditions. Finally, we can get the idea tree from the topic citation network by keeping the most important citation.\\

\subsection*{The calculation of knowledge entropy}
\noindent Based on the idea tree, we can start from the structure of the tree and utilize structural information to measure the knowledge quality of academic articles. In particular, we quantify this knowledge quality in the form of knowledge entropy.\\

\noindent For an article $a$, we are inspired by previous work\cite{li2016structural} and define the subtree entropy of $a$ as follows:\\
\begin{equation}
H\left(a\right)=-\frac{g_{a}}{2m}\log\left(\frac{V_{a}}{V_{a^-}}\right)
\tag{M5}
\end{equation}
\noindent where $g_{a}$ represents the number of the cut edges between the nodes in the subtree rooted on $a$ in the idea tree and the nodes out of the subtree. $V_{a}$ represents the number of the nodes in subtree rooted on $a$ while $V_{a^-}$ represents the number of the nodes in subtree rooted on $a$'s parent node. Generally speaking, it satisifies the relationship that $V_{a^-} \geq V_{a}+1$.\\

\noindent With the definition of subtree entropy above, the definition of mutual knowledge entropy is also given as follows:\\
\begin{equation}
I\left(a,b\right)=-\frac{g_{ab}}{4m}\log\left(\frac{V_{a}V_{b}}{V_{{ab}^-}^{2}}\right)
\tag{M6}
\end{equation}
\noindent where $g_{ab}$ represents the number of the cut edges between the nodes in the subtree rooted on $a$ and the subtree rooted on $b$ in the idea tree acquired above and the nodes out of the two subtree. $V_{a}$ represents the number of the nodes in subtree rooted on $a$, $V_{b}$ represents the number of the nodes in subtree rooted on $b$, while $V_{{ab}^-}$ represents the number of the nodes in subtree rooted on $a$ and $b$'s parent node, which indicates that $a$ and $b$ should have the same parent node, or the two nodes should locate in similar positions in the idea tree.\\

\noindent Based on the subtree entropy and mutual knowledge entropy above, the definition of knowledge entropy is given as follows: at timestamp $t$, the topic citation network is $G^t=\left(V^t,E^t\right)$. The idea tree extracted from the network is $IdeaTree\left(G^t\right)$. For any paper $v$ belong to $IdeaTree\left(G^t\right)$ except the seminal work, we define its knowledge entropy ${KE}^t\left(v\right)$ as:
\begin{equation}
  {KE}^t\left(v\right)=\left|H^t\left(v\right)-\sum_{v_i\in C^t\left(v\right)}{H^t\left(v_i\right)}+\sum_{v_i,v_j\in C^t\left(v\right),i\neq j}{I^t\left(v_i,v_j\right)}\right|  
  \tag{M7}
\end{equation}
\noindent where $H^t\left(v\right)$ represents the subtree entropy of the subtree led by node $v$ at $t$, $C^t\left(v\right)$ represents the children node of $v$ in $IdeaTree\left(G^t\right)$ at $t$, and $I^t\left(v_i,v_j\right)$ represents the mutual knowledge entropy of the subtree led by $v_i$ and $v_j$ at $t$.\\

\noindent The equation above can be illustrated that the knowledge entropy of Node $v$ is the subtree entropy of $v$ minus all the subtree entropy of $v$'s children, while the mutual knowledge entropy of the children is compensated. To put it another way, the knowledge entropy is $v$'s total influence on the research area minus all $v$'s children's total influence, which describes $v$'s quality of knowledge or novelty itself.\\

\subsection*{Fitting of idea limit formula}
\noindent Assuming that the form of the idea limit formula is ${\Delta D}^t(v) = \log \frac{KE^{t}(v)}{\left(t-t_{0}\right)^{\gamma}}$, with the help of the least squares method, we use the evolution data of idea tree and knowledge entropy as the sample data to fit the time attenuation coefficient $\gamma$. For any high knowledge entropy node $v$ in any idea tree, node $v$ becomes visible at $t_0$, i.e. $KE^{t_0}\left(v\right)\geq M$, taking node $v$ as a reference, the visible depth of the subtree led by node $v$ is 0 at $t_0$. We know that the maximum visible depth $MaxVD_{subtree_v}$ of the subtree led by node $v$ up to the current time $t_{now}$ is $VD^{t_{now}}_{subtree_v}$, at any moment $\overline{t}$ between $t_0$ and $t_{now}$, we can get the sample data ${\Delta D}^{\overline{t}}(v)$ of ${\Delta D}^t(v)$: ${\Delta D}^{\overline{t}}(v)=MaxVD_{subtree_v}-VD^{\overline{t}}_{subtree_v}$. Similarly, we know that the knowledge entropy of node $v$ at $\overline{t}$ is $KE^{\overline{t}}(v)$, therefore, we can get the sample data $(KE^{\overline{t}_i}(v_j), \overline{t}_i-t_0, {\Delta D}^{\overline{t}_i}(v_j))$ from all the idea trees to fit the formula. We transform the form of the formula to $\log KE^{t}(v)-{\Delta D}^t(v)=\gamma \log (t-t_0)$, and the value of a can be obtained according to the least squares method:\\

\begin{equation}
  \hat{\gamma}=\mathop{\arg\min}_{\gamma}\sum _{i,j}((\log KE^{\overline{t}_i}(v_j)-{\Delta D}^{\overline{t}_i}(v_j))-\gamma \log (\overline{t}_i-t_0))^2
  \tag{M8}
\end{equation}

\noindent Let the derivative of the objective function to $\gamma$ be 0, we can get $\hat{\gamma}=\frac{\sum _{i,j}(\log KE^{\overline{t}_i}(v_j)-{\Delta D}^{\overline{t}_i}(v_j))}{\sum _{i,j}\log (\overline{t}_i-t_0)}$, and the fitting result of  $\gamma$ is 1.914.

\section*{Supplementary Materials - Scientific X-ray}
\section*{S1 DATA DETAILS}
\noindent We collect and integrate academic data from bibliographic databases, including but not limit to IEEE, ACM, arXiv, Elsevier, and Spring. We then select all 71 431 high-impact publications with more than 1000 citations as pioneering works and construct topic citation networks for them. All leading articles were published between 1800 and 2021, and their research interests cover 294 fields in 16 disciplines: History, Computer science, Environmental science, Geology, Psychology, Mathematics, Physics, Materials science, Philosophy, Biology, Medicine, Sociology, Art, Economics, Chemistry, and Political science. Data details of the leading articles and topic overviews appearing in each section of main text and supplementary materials are shown as follows.\\
\subsection*{S1.1 The framework of Scientific X-ray}
\noindent Details of leading articles are listed in Table \ref{tab_s1_1}, and topic overviews are listed in Table \ref{tab_s1_2}.\\
\setcounter{table}{0}

%定义编号格式，在数字序号前加字符“A"

\renewcommand{\thetable}{S1-\arabic{table}}
\begin{table}[H]
    \centering
\begin{tabular}{p{8.5cm}p{1cm}p{5cm}}
\toprule
                                                            Leading article &  Year &           Journal/Conference Series \\
\midrule
                                        The End of History and the Last Man &  1992 &                                    Free Press \\
 A New Statistical Method for Haplotype Reconstruction from Population Data &  2001 &  American Journal of Human Genetics \\
\bottomrule
\end{tabular}
\captionsetup{labelfont=bf}
        \caption{Leading article details}
		\label{tab_s1_1} 
\end{table}

\begin{table}[H]
    \centering
\begin{tabular}{p{12.5cm}p{1cm}p{1cm}}
\toprule
                                                            Leading article &  Nodes &  Edges \\
\midrule
                                        The End of History and the Last Man &   6604 &   9042 \\
 A New Statistical Method for Haplotype Reconstruction from Population Data &   6759 &  27569 \\
\bottomrule
\end{tabular}
\captionsetup{labelfont=bf}
        \caption{Topic overviews}
		\label{tab_s1_2} 
\end{table}

\subsection*{S1.2 The evolution patterns of scientific topics}
\noindent Details of leading articles are listed in Table \ref{tab_s1_3}, and topic overviews are listed in Table \ref{tab_s1_4}.\\

\begin{table}[H]
    \centering
\begin{tabular}{p{8.5cm}p{1cm}p{5cm}}
\toprule
                                                                              Leading article &  Year &                                                       Journal/Conference Series \\
\midrule
                                                          Geographic Information Systems and Science &  2001 &                   Wiley\\
                                 SphereFace: Deep Hypersphere Embedding for Face Recognition &  2017 &                                   Computer Vision and Pattern Recognition(CVPR) \\
                            Range Shifts and Adaptive Responses to Quaternary Climate Change &  2001 &                                                                         Science \\
                                                    DEEP CONTEXTUALIZED WORD REPRESENTATIONS &  2018 &  North American Chapter of the Association for Computational Linguistics(NAACL) \\
 DoReFa-Net: Training Low Bitwidth Convolutional Neural Networks with Low Bitwidth Gradients &  2016 &                                        arXiv: Neural and Evolutionary Computing \\
           Evolution of the Altaid tectonic collage and Palaeozoic crustal growth in Eurasia &  1993 &                                                                          Nature \\
\bottomrule
\end{tabular}

\captionsetup{labelfont=bf}
        \caption{Leading article details}
		\label{tab_s1_3} 
\end{table}

\begin{table}[H]
    \centering
\begin{tabular}{p{12.5cm}p{1cm}p{1cm}}
\toprule
                                                                              Leading article &  Nodes &  Edges \\
\midrule
                                                          Geographic Information Systems and Science &   1744 &   2619 \\
                                 SphereFace: Deep Hypersphere Embedding for Face Recognition &   1146 &   4488 \\
                            Range Shifts and Adaptive Responses to Quaternary Climate Change &   1808 &   8561 \\
                                                    DEEP CONTEXTUALIZED WORD REPRESENTATIONS &   5215 &  26245 \\
 DoReFa-Net: Training Low Bitwidth Convolutional Neural Networks with Low Bitwidth Gradients &   1152 &   7485 \\
           Evolution of the Altaid tectonic collage and Palaeozoic crustal growth in Eurasia &   2552 &  53626 \\
\bottomrule
\end{tabular}
\captionsetup{labelfont=bf}
        \caption{Topic overviews}
		\label{tab_s1_4} 
\end{table}

\subsection*{S1.3 The parse of science’s 2015 breakthrough}
\noindent Details of leading articles are listed in Table \ref{tab_s1_5}, and topic overviews are listed in Table \ref{tab_s1_6}.\\

\begin{table}[H]
    \centering
\begin{tabular}{p{9cm}p{1cm}p{4.5cm}}
\toprule
                                                                                                                                                         Leading article &  Year & Journal/Conference Series \\
\midrule
                                                                                         A programmable dual-RNA-guided DNA endonuclease in adaptive bacterial immunity. &  2012 &                   Science \\
                                                                                                  The Pluto system: Initial results from its exploration by New Horizons &  2015 &                   Science \\
                                                                                                                          The ancestry and affiliations of Kennewick Man &  2015 &                    Nature \\
                                                                                                                            Many psychology papers fail replication test &  2015 &                   Science \\
                                                                                    Homo naledi, a new species of the genus Homo from the Dinaledi Chamber, South Africa &  2015 &                     eLife \\
                                                                                           Broad plumes rooted at the base of the Earth's mantle beneath major hotspots. &  2015 &                    Nature \\
 Efficacy and effectiveness of an rVSV-vectored vaccine expressing Ebola surface glycoprotein: interim results from the Guinea ring vaccination cluster-randomised trial &  2015 &                The Lancet \\
                                                                                       Complete absence of thebaine biosynthesis under home-brew fermentation conditions &  2015 &                   bioRxiv \\
                                                                                         Structural and functional features of central nervous system lymphatic vessels. &  2015 &                    Nature \\
                                                                                Loophole-free Bell inequality violation using electron spins separated by 1.3 kilometres &  2015 &                    Nature \\
\bottomrule
\end{tabular}

\captionsetup{labelfont=bf}
        \caption{Leading article details}
		\label{tab_s1_5} 
\end{table}

\begin{table}[H]
    \centering
    
\begin{tabular}{p{12.5cm}p{1cm}p{1cm}}
\toprule
                                                                                                                                                         Leading article &  Nodes &   Edges \\
\midrule
                                                                                         A programmable dual-RNA-guided DNA endonuclease in adaptive bacterial immunity. &   7784 &  134352 \\
                                                                                                  The Pluto system: Initial results from its exploration by New Horizons &    312 &    2151 \\
                                                                                                                          The ancestry and affiliations of Kennewick Man &    143 &     458 \\
                                                                                                                            Many psychology papers fail replication test &     73 &      83 \\
                                                                                    Homo naledi, a new species of the genus Homo from the Dinaledi Chamber, South Africa &    426 &    1255 \\
                                                                                           Broad plumes rooted at the base of the Earth's mantle beneath major hotspots. &    410 &    1330 \\
 Efficacy and effectiveness of an rVSV-vectored vaccine expressing Ebola surface glycoprotein: interim results from the Guinea ring vaccination cluster-randomised trial &    577 &    2240 \\
                                                                                       Complete absence of thebaine biosynthesis under home-brew fermentation conditions &      5 &       5 \\
                                                                                         Structural and functional features of central nervous system lymphatic vessels. &   1871 &    7883 \\
                                                                                Loophole-free Bell inequality violation using electron spins separated by 1.3 kilometres &   1649 &    7917 \\
\bottomrule
\end{tabular}

\captionsetup{labelfont=bf}
        \caption{Topic overviews}
		\label{tab_s1_6} 
\end{table}

\subsection*{S1.4 Examples of six evolution patterns}
\noindent Details of leading articles are listed in Table \ref{tab_s1_7}, \ref{tab_s1_8}, and topic overviews are listed in Table \ref{tab_s1_9}.\\

\begin{table}[H]
    \centering
\begin{tabular}{p{8.5cm}p{1cm}p{5cm}}
\toprule
                                                                   Leading article &  Year &                                                        Journal/Conference Series \\
\midrule
                   Principal component analysis: a review and recent developments. &  2016 &                                Philosophical Transactions of the Royal Society A \\
                                         Multilevel Analysis : Techniques and Applications, Third Edition &  2017 & Routledge \\
                                                                    An Introduction to Medical Statistics &  1987 &   Oxford University Press\\
                                                    TensorFlow: a system for large-scale machine learning &  2016 &                                Operating Systems Design and Implementation(OSDI) \\
                                             Recent Arctic amplification and extreme mid-latitude weather &  2014 &                                                                Nature Geoscience \\
 Processing seismic ambient noise data to obtain reliable broad-band surface wave dispersion measurements &  2007 &                                                Geophysical Journal International \\
                                                    Overcoming catastrophic forgetting in neural networks &  2017 &  Proceedings of the National Academy of Sciences of the United States of America \\
                                                         Globally and locally consistent image completion &  2017 &                                                     ACM Transactions on Graphics \\
                                                                    Improved techniques for training GANs &  2016 &                                   Neural Information Processing Systems(NeurIPS) \\
                                                                  Matching networks for one shot learning &  2016 &                                   Neural Information Processing Systems(NeurIPS) \\
                                                           A Meta-Analysis of Global Urban Land Expansion &  2011 &                                                                         PLOS ONE \\
                                                                              Cleavage of GSDMD by inflammatory caspases determines pyroptotic cell death &  2015 &                                                                           Nature \\
                              
\bottomrule
\end{tabular}

\captionsetup{labelfont=bf}
        \caption{Leading article details}
		\label{tab_s1_7} 
\end{table}

\begin{table}[H]
    \centering
\begin{tabular}{p{8.5cm}p{1cm}p{5cm}}
\toprule
                                                                   Leading article &  Year &                                                        Journal/Conference Series \\
\midrule

                                                                  The Kinetics Human Action Video Dataset &  2017 &                                   arXiv: Computer Vision and Pattern Recognition \\
                          Estimating Corn Leaf Chlorophyll Concentration from Leaf and Canopy Reflectance &  2000 &                                                    Remote Sensing of Environment \\
                                                                 Thermal remote sensing of urban climates &  2003 &                                                    Remote Sensing of Environment \\
           Weyl Semimetal Phase in Noncentrosymmetric Transition-Metal Monophosphides &  2015 &                                                                Physical Review X \\
       Non-ideal interactions in calcic amphiboles and their bearing on amphibole-plagioclase thermometry &  1994 &                                        Contributions to Mineralogy and Petrology \\
                 A dynamic global vegetation model for studies of the coupled atmosphere‐biosphere system &  2005 &                                                     Global Biogeochemical Cycles \\
                                             Robust Responses of the Hydrological Cycle to Global Warming &  2006 &                                                               Journal of Climate \\
                                                              Prototypical Networks for Few-shot Learning &  2017 &                                   Neural Information Processing Systems(NeurIPS) \\
             Unsupervised Representation Learning with Deep Convolutional Generative Adversarial Networks &  2016 &                       International Conference on Learning Representations(ICLR) \\
                                                            Feature Pyramid Networks for Object Detection &  2017 &                                    Computer Vision and Pattern Recognition(CVPR) \\
                        Inception-v4, Inception-ResNet and the Impact of Residual Connections on Learning &  2016 &                             National Conference on Artificial Intelligence(AAAI) \\
                                  High Serum IgG4 Concentrations in Patients with Sclerosing Pancreatitis &  2001 &                                              The New England Journal of Medicine \\
\bottomrule
\end{tabular}

\captionsetup{labelfont=bf}
        \caption{Leading article details}
		\label{tab_s1_8} 
\end{table}

\begin{table}[H]
    \centering
\begin{tabular}{p{12.5cm}p{1cm}p{1cm}}
\toprule
                                                                   Leading article &  Nodes &  Edges \\
\midrule
                   Principal component analysis: a review and recent developments. &   1430 &   1545 \\
                                         Multilevel Analysis : Techniques and Applications, Third Edition &   2322 &   3268 \\
                                                                    An Introduction to Medical Statistics &   2180 &   2730 \\
                                                    TensorFlow: a system for large-scale machine learning &   5941 &   9676 \\
                                             Recent Arctic amplification and extreme mid-latitude weather &   1037 &   6187 \\
 Processing seismic ambient noise data to obtain reliable broad-band surface wave dispersion measurements &   1308 &  10856 \\
                                                    Overcoming catastrophic forgetting in neural networks &   1648 &  10101 \\
                                                         Globally and locally consistent image completion &   1042 &   4747 \\
                                                                    Improved techniques for training GANs &   2394 &  17336 \\
                                                                  Matching networks for one shot learning &   1795 &   9672 \\
                                                           A Meta-Analysis of Global Urban Land Expansion &   1119 &   3031 \\
                              Cleavage of GSDMD by inflammatory caspases determines pyroptotic cell death &   1584 &  14187 \\
                                                                  The Kinetics Human Action Video Dataset &   1202 &   6373 \\
                          Estimating Corn Leaf Chlorophyll Concentration from Leaf and Canopy Reflectance &   1282 &   7440 \\
                                                                 Thermal remote sensing of urban climates &   1540 &  12323 \\
           Weyl Semimetal Phase in Noncentrosymmetric Transition-Metal Monophosphides &   1162 &  14237 \\
       Non-ideal interactions in calcic amphiboles and their bearing on amphibole-plagioclase thermometry &   1661 &   7493 \\
                 A dynamic global vegetation model for studies of the coupled atmosphere‐biosphere system &   1522 &  10236 \\
                                             Robust Responses of the Hydrological Cycle to Global Warming &   3151 &  24511 \\
                                                              Prototypical Networks for Few-shot Learning &   2310 &  14410 \\
             Unsupervised Representation Learning with Deep Convolutional Generative Adversarial Networks &   6492 &  50667 \\
                                                            Feature Pyramid Networks for Object Detection &   5896 &  36669 \\
                        Inception-v4, Inception-ResNet and the Impact of Residual Connections on Learning &   3551 &  11192 \\
                                  High Serum IgG4 Concentrations in Patients with Sclerosing Pancreatitis &   2056 &  33255 \\
\bottomrule
\end{tabular}
\captionsetup{labelfont=bf}
        \caption{Topic overviews}
		\label{tab_s1_9} 
\end{table}

\subsection*{S1.5 Measuring the development potential of specific topics}
\subsubsection*{Geoscience}
\noindent Details of leading articles are listed in Table \ref{tab_s1_10}, and topic overviews are listed in Table \ref{tab_s1_11}.\\
\begin{table}[H]
    \centering
\begin{tabular}{p{8.5cm}p{1cm}p{5cm}}
\toprule
                                                                   Leading article &  Year &                                                        Journal/Conference Series \\
\midrule

                                                                                                                                                                   The effect of human mobility and control measures on the COVID-19 epidemic in China &  2020 &                                                                          Science \\
                                                                                                                                                                                           Mangroves among the most carbon-rich forests in the tropics &  2011 &                                                                Nature Geoscience \\
 Global, regional, and national comparative risk assessment of 79 behavioural, environmental and occupational, and metabolic risks or clusters of risks in 188 countries, 1990–2013: a systematic analysis for the Global Burden of Disease Study 2013 &  2015 & The Lancet\\
                                                                                                                                                                         Global land use change, economic globalization, and the looming land scarcity &  2011 &  Proceedings of the National Academy of Sciences of the United States of America \\
                                                                                                                                      Hemispheric and large‐scale land‐surface air temperature variations: An extensive revision and an update to 2010 &  2012 &                                                  Journal of Geophysical Research \\
                                                                                                                                                                                                      Linear Mixed-Effects Models using 'Eigen' and S4 &  2015 &  Journal of Statistical Software\\
                                                                                                                                                                                                             The Transiting Exoplanet Survey Satellite &  2014 &                                          arXiv: Earth and Planetary Astrophysics \\
                                                                                                                                                        'Structure-from-Motion' photogrammetry: A low-cost, effective tool for geoscience applications &  2012 &                                                                    Geomorphology \\
                                                                                                                                                                                      Object-based cloud and cloud shadow detection in Landsat imagery &  2012 &                                                    Remote Sensing of Environment \\
                                                                                                                                                                              Bedmap2: improved ice bed, surface and thickness datasets for Antarctica &  2012 &                                                                   The Cryosphere \\
\bottomrule
\end{tabular}

\captionsetup{labelfont=bf}
        \caption{Leading article details}
		\label{tab_s1_10} 
\end{table}

\begin{table}[H]
    \centering
\begin{tabular}{p{12.5cm}p{1cm}p{1cm}}
\toprule
                                                                   Leading article &  Nodes &  Edges \\
\midrule
                  The effect of human mobility and control measures on the COVID-19 epidemic in China &   1024 &   2782 \\
                                                                                                                                                                                           Mangroves among the most carbon-rich forests in the tropics &   1283 &   9708 \\
 Global, regional, and national comparative risk assessment of 79 behavioural, environmental and occupational, and metabolic risks or clusters of risks in 188 countries, 1990–2013: a systematic analysis for the Global Burden of Disease Study 2013 &   1430 &   4036 \\
                                                                                                                                                                         Global land use change, economic globalization, and the looming land scarcity &   1647 &   5552 \\
                                                                                                                                      Hemispheric and large‐scale land‐surface air temperature variations: An extensive revision and an update to 2010 &   1537 &   5990 \\
                                                                                                                                                                                                      Linear Mixed-Effects Models using 'Eigen' and S4 &   5945 &   9040 \\
                                                                                                                                                                                                             The Transiting Exoplanet Survey Satellite &   1464 &   9277 \\
                                                                                                                                                        'Structure-from-Motion' photogrammetry: A low-cost, effective tool for geoscience applications &   1747 &   8570 \\
                                                                                                                                                                                      Object-based cloud and cloud shadow detection in Landsat imagery &   1072 &   5312 \\
                                                                                                                                                                              Bedmap2: improved ice bed, surface and thickness datasets for Antarctica &   1207 &   9363 \\
\bottomrule
\end{tabular}
\captionsetup{labelfont=bf}
        \caption{Topic overviews}
		\label{tab_s1_11} 
\end{table}

\subsubsection*{Deep learning}

\noindent Details of leading articles are listed in Table \ref{tab_s1_12}, and topic overviews are listed in Table \ref{tab_s1_13}.\\

\begin{table}[H]
    \centering
\begin{tabular}{p{8.5cm}p{1cm}p{5cm}}
\toprule
                                                                   Leading article &  Year &                                                        Journal/Conference Series \\
\midrule

                                                               Attention is All you Need &  2017 &                                   Neural Information Processing Systems(NeurIPS) \\
                                         Prototypical Networks for Few-shot Learning &  2017 &                                   Neural Information Processing Systems(NeurIPS) \\
                                             Matching networks for one shot learning &  2016 &                                   Neural Information Processing Systems(NeurIPS) \\
                    Semi-Supervised Classification with Graph Convolutional Networks &  2017 &                       International Conference on Learning Representations(ICLR) \\
                      Understanding deep learning requires rethinking generalization &  2017 &                       International Conference on Learning Representations(ICLR) \\
           A Style-Based Generator Architecture for Generative Adversarial Networks. &  2018 &                                    Computer Vision and Pattern Recognition(CVPR) \\
                                                 Universal Adversarial Perturbations &  2017 &                                    Computer Vision and Pattern Recognition(CVPR) \\
 Distillation as a Defense to Adversarial Perturbations Against Deep Neural Networks &  2016 &                                      IEEE Symposium on Security and Privacy(S\&P) \\
                 DeepFool: A Simple and Accurate Method to Fool Deep Neural Networks &  2016 &                                    Computer Vision and Pattern Recognition(CVPR) \\
                               Overcoming catastrophic forgetting in neural networks &  2017 &  Proceedings of the National Academy of Sciences of the United States of America \\
\bottomrule
\end{tabular}

\captionsetup{labelfont=bf}
        \caption{Leading article details}
		\label{tab_s1_12} 
\end{table}

\begin{table}[H]
    \centering
\begin{tabular}{p{12.5cm}p{1cm}p{1cm}}
\toprule
                                                                   Leading article &  Nodes &  Edges \\
\midrule
                                                           Attention is All you Need &  15522 &  83563 \\
                                         Prototypical Networks for Few-shot Learning &   2310 &  14410 \\
                                             Matching networks for one shot learning &   1795 &   9672 \\
                    Semi-Supervised Classification with Graph Convolutional Networks &   3966 &  15464 \\
                      Understanding deep learning requires rethinking generalization &   1578 &   7233 \\
           A Style-Based Generator Architecture for Generative Adversarial Networks. &   1837 &   6715 \\
                                                 Universal Adversarial Perturbations &   1236 &   7888 \\
 Distillation as a Defense to Adversarial Perturbations Against Deep Neural Networks &   1531 &  11567 \\
                 DeepFool: A Simple and Accurate Method to Fool Deep Neural Networks &   2328 &  18905 \\
                               Overcoming catastrophic forgetting in neural networks &   1648 &  10101 \\   
\bottomrule
\end{tabular}
\captionsetup{labelfont=bf}
        \caption{Topic overviews}
		\label{tab_s1_13} 
\end{table}

\subsubsection*{Computer vision}

\noindent Details of leading articles are listed in Table \ref{tab_s1_14}, and topic overviews are listed in Table \ref{tab_s1_15}.\\

\begin{table}[H]
    \centering
\begin{tabular}{p{8.5cm}p{1cm}p{5cm}}
\toprule
                                                                   Leading article &  Year &                                                        Journal/Conference Series \\
\midrule
                                         Rethinking Atrous Convolution for Semantic Image Segmentation &  2017 &                  arXiv: Computer Vision and Pattern Recognition \\
                        PointNet++: Deep Hierarchical Feature Learning on Point Sets in a Metric Space &  2017 &                  Neural Information Processing Systems(NeurIPS) \\
 The SYNTHIA Dataset: A Large Collection of Synthetic Images for Semantic Segmentation of Urban Scenes &  2016 &                   Computer Vision and Pattern Recognition(CVPR) \\
                                       ArcFace: Additive Angular Margin Loss for Deep Face Recognition &  2018 &                   Computer Vision and Pattern Recognition(CVPR) \\
                                                                    YOLO9000: Better, Faster, Stronger &  2017 &                   Computer Vision and Pattern Recognition(CVPR) \\
                                               You Only Look Once: Unified, Real-Time Object Detection &  2016 &                   Computer Vision and Pattern Recognition(CVPR) \\
                                                                     Deformable Convolutional Networks &  2017 &               International Conference on Computer Vision(ICCV) \\
                                              Image Super-Resolution Using Deep Convolutional Networks &  2016 &  IEEE Transactions on Pattern Analysis and Machine Intelligence \\
                                                        Dynamic Graph CNN for Learning on Point Clouds &  2019 &                                    ACM Transactions on Graphics \\
                                      Image-to-Image Translation with Conditional Adversarial Networks &  2017 &                   Computer Vision and Pattern Recognition(CVPR) \\
\bottomrule
\end{tabular}

\captionsetup{labelfont=bf}
        \caption{Leading article details}
		\label{tab_s1_14} 
\end{table}

\begin{table}[H]
    \centering
\begin{tabular}{p{12.5cm}p{1cm}p{1cm}}
\toprule
                                                                   Leading article &  Nodes &  Edges \\
\midrule
                                         Rethinking Atrous Convolution for Semantic Image Segmentation &   2419 &   9098 \\
                        PointNet++: Deep Hierarchical Feature Learning on Point Sets in a Metric Space &   2563 &  20653 \\
 The SYNTHIA Dataset: A Large Collection of Synthetic Images for Semantic Segmentation of Urban Scenes &   1131 &   6776 \\
                                       ArcFace: Additive Angular Margin Loss for Deep Face Recognition &   1223 &   3491 \\
                                                                    YOLO9000: Better, Faster, Stronger &   5027 &  17497 \\
                                               You Only Look Once: Unified, Real-Time Object Detection &   9007 &  40418 \\
                                                                     Deformable Convolutional Networks &   1582 &   6907 \\
                                              Image Super-Resolution Using Deep Convolutional Networks &   3483 &  20884 \\
                                                        Dynamic Graph CNN for Learning on Point Clouds &   1136 &   4954 \\
                                      Image-to-Image Translation with Conditional Adversarial Networks &   7456 &  38130 \\
\bottomrule
\end{tabular}
\captionsetup{labelfont=bf}
        \caption{Topic overviews}
		\label{tab_s1_15} 
\end{table}

\subsubsection*{Natural language processing}

\noindent Details of leading articles are listed in Table \ref{tab_s1_16}, and topic overviews are listed in Table \ref{tab_s1_17}.\\

\begin{table}[H]
    \centering
\begin{tabular}{p{8.5cm}p{1cm}p{5cm}}
\toprule
                                                                   Leading article &  Year &                                                        Journal/Conference Series \\
\midrule
                     A BROAD-COVERAGE CHALLENGE CORPUS FOR SENTENCE UNDERSTANDING THROUGH INFERENCE &  2018 &  North American Chapter of the Association for Computational Linguistics(NAACL) \\
                                    Get To The Point: Summarization with Pointer-Generator Networks &  2017 &                   Meeting of the Association for Computational Linguistics(ACL) \\
                                                    Enriching Word Vectors with Subword Information &  2017 &                   Transactions of the Association for Computational Linguistics \\
                                        SQuAD: 100,000+ Questions for Machine Comprehension of Text &  2016 &                         Empirical Methods in Natural Language Processing(EMNLP) \\
  Listen, attend and spell: A neural network for large vocabulary conversational speech recognition &  2016 &    International Conference on Acoustics, Speech, and Signal Processing(ICASSP) \\
 Google's Neural Machine Translation System: Bridging the Gap between Human and Machine Translation &  2016 &                                                 arXiv: Computation and Language \\
                                        Neural Machine Translation of Rare Words with Subword Units &  2016 &                   Meeting of the Association for Computational Linguistics(ACL) \\
                                  Improving Neural Machine Translation Models with Monolingual Data &  2016 &                   Meeting of the Association for Computational Linguistics(ACL) \\
                   BERT: Pre-training of Deep Bidirectional Transformers for Language Understanding &  2018 &                                                 arXiv: Computation and Language \\
                           XLNet: Generalized Autoregressive Pretraining for Language Understanding &  2019 &                                                 arXiv: Computation and Language \\
\bottomrule
\end{tabular}

\captionsetup{labelfont=bf}
        \caption{Leading article details}
		\label{tab_s1_16} 
\end{table}

\begin{table}[H]
    \centering
\begin{tabular}{p{12.5cm}p{1cm}p{1cm}}
\toprule
                                                                   Leading article &  Nodes &  Edges \\
\midrule
                     A BROAD-COVERAGE CHALLENGE CORPUS FOR SENTENCE UNDERSTANDING THROUGH INFERENCE &   1220 &   7839 \\
                                    Get To The Point: Summarization with Pointer-Generator Networks &   1863 &   9758 \\
                                                    Enriching Word Vectors with Subword Information &   4451 &  12017 \\
                                        SQuAD: 100,000+ Questions for Machine Comprehension of Text &   1160 &   8290 \\
  Listen, attend and spell: A neural network for large vocabulary conversational speech recognition &   1231 &   4746 \\
 Google's Neural Machine Translation System: Bridging the Gap between Human and Machine Translation &   3275 &  11551 \\
                                        Neural Machine Translation of Rare Words with Subword Units &   3749 &  25844 \\
                                  Improving Neural Machine Translation Models with Monolingual Data &   1306 &   7073 \\
                   BERT: Pre-training of Deep Bidirectional Transformers for Language Understanding &   6898 &  20124 \\
                           XLNet: Generalized Autoregressive Pretraining for Language Understanding &   1519 &   4564 \\
\bottomrule
\end{tabular}
\captionsetup{labelfont=bf}
        \caption{Topic overviews}
		\label{tab_s1_17} 
\end{table}

\subsubsection*{Data mining}

\noindent Details of leading articles are listed in Table \ref{tab_s1_18}, and topic overviews are listed in Table \ref{tab_s1_19}.\\

\begin{table}[H]
    \centering
\begin{tabular}{p{8.5cm}p{1cm}p{5cm}}
\toprule
                                                                   Leading article &  Year &                                                        Journal/Conference Series \\
\midrule
            Inductive Representation Learning on Large Graphs &  2017 &              Neural Information Processing Systems(NeurIPS) \\
                 Wide \& Deep Learning for Recommender Systems &  2016 &                   Conference on Recommender Systems(RecSys) \\
 Membership Inference Attacks Against Machine Learning Models &  2017 &                 IEEE Symposium on Security and Privacy(S\&P) \\
             Deep Neural Networks for YouTube Recommendations &  2016 &                   Conference on Recommender Systems(RecSys) \\
                               Neural Collaborative Filtering &  2017 &                                     The Web Conference(WWW) \\
                      Deep Learning with Differential Privacy &  2016 &                   Computer and Communications Security(CCS) \\
                                     Graph Attention Networks &  2018 &  International Conference on Learning Representations(ICLR) \\
   Modeling Relational Data with Graph Convolutional Networks &  2018 &                      European Semantic Web Conference(ESWC) \\
                     XGBoost: A Scalable Tree Boosting System &  2016 &                    Knowledge Discovery and Data Mining(KDD) \\
                            Structural Deep Network Embedding &  2016 &                    Knowledge Discovery and Data Mining(KDD) \\
\bottomrule
\end{tabular}

\captionsetup{labelfont=bf}
        \caption{Leading article details}
		\label{tab_s1_18} 
\end{table}

\begin{table}[H]
    \centering
\begin{tabular}{p{12.5cm}p{1cm}p{1cm}}
\toprule
                                                                   Leading article &  Nodes &  Edges \\
\midrule
            Inductive Representation Learning on Large Graphs &   2071 &  11398 \\
                 Wide \& Deep Learning for Recommender Systems &   1134 &   4563 \\
 Membership Inference Attacks Against Machine Learning Models &   1064 &   6158 \\
             Deep Neural Networks for YouTube Recommendations &   1161 &   3799 \\
                               Neural Collaborative Filtering &   1711 &   8466 \\
                      Deep Learning with Differential Privacy &   1548 &   8756 \\
                                     Graph Attention Networks &   1101 &   3443 \\
   Modeling Relational Data with Graph Convolutional Networks &   1103 &   3796 \\
                     XGBoost: A Scalable Tree Boosting System &   6884 &  12437 \\
                            Structural Deep Network Embedding &   1417 &   7747 \\
\bottomrule
\end{tabular}
\captionsetup{labelfont=bf}
        \caption{Topic overviews}
		\label{tab_s1_19} 
\end{table}

\section*{S2 DERIVATION DETAILS}
\subsection*{S2.1 Idea tree extraction algorithm}

\noindent There exists three steps in the algorithm. Initially, distance between nodes in high-dimensional space is measured via reference relationships. Then with the encoding method of random walk, reduction index is acquired by measuring the weighted distance among the papers in the network, which are then used to select edges to cut. Specifically, we will cut the edges between the node pairs which have largest reduction index difference and then finish the turning from the topic citation network to idea tree.\\

\subsubsection*{The distance of academic articles in high-dimensional space}

\noindent The first step is to measure the distance between the nodes by citation relationships. Particularly, we utilize graph embedding to measure such distance in high dimensional vector space. As for any scientific topic, we construct topic citation network $G\left(V,E\right)$, among which the $V$ represents the set of all the nodes and $E$ represents the set of all the edges. Specially, $n=\left\Vert V\right\Vert $ is defined as the amount of the nodes in the network while $m=\left\Vert E\right\Vert $ is defined as the amount of the edges. $A$ represents the adjacency matrix of the network, with the form of:\\
$$
A=\left(\begin{array}{cccc}
A_{11} & A_{12} & \ldots & A_{1n}\\
A_{21} & A_{22} & \ldots & A_{2n}\\
\vdots & \vdots & \ddots & \vdots\\
A_{n1} & A_{n2} & \ldots & A_{nn}
\end{array}\right)
$$
\noindent where $A_{ij}=1$ represents that there exists reference relationships that paper $v_i$ cites paper $v_j$. However, due to the existence of errors in the real data, $A_{ij}=0$ doesn't absolutely means that paper $v_i$ doesn't cite paper $v_j$. Actually, there seldom happens the phenomenon that two papers cite each other, which is determined by the reason that papers are usually published in sequential. Whereas, we do find few data in such form in the database, which we then turn such cite-each-other relationship to normal reference relationship following the rule that the paper published later follows the paper published priorly.\\

\noindent Especially, we noticed that the pioneering paper cites none of the papers in the network since it's the earliest one, which blocks our later calculation for eigenvalues and eigenvectors. Considering this, we involve self-citation or self-loop to the pioneering paper, which allows subsequent eigenvalue decomposition. After the process above, we get $W$, the adjacency matrix with self-loop:\\
$$
W=\left(\begin{array}{cccc}
W_{11} & W_{12} & \ldots & W_{1n}\\
W_{21} & W_{22} & \ldots & W_{2n}\\
\vdots & \vdots & \ddots & \vdots\\
W_{n1} & W_{n2} & \ldots & W_{nn}
\end{array}\right)
$$
\noindent where $W_{ij}=A_{ij}$ when $v_i \neq v_j$, $W_{ij}=0$ when $v_i=v_j$ and $\underset{j,j\neq i}{\sum}A_{ij}>0$, and $W_{ij}=1$ when $v_i=v_j$ and $\underset{j,j\neq i}{\sum}A_{ij}=0$.
Continue to process the adjacency matrix with self-loop $W$, we get the output matrix with self loop $D$:\\
$$
D=\left(\begin{array}{cccc}
d_{1} & 0 & \ldots & 0\\
0 & d_{2} & \ldots & 0\\
\vdots & \vdots & \ddots & \vdots\\
0 & 0 & \ldots & d_{n}
\end{array}\right)
$$
\noindent which is a diagonal matrix with $d_{i}=\underset{j}{\sum}W_{ij}$ and $d_{i} \neq 0$.\\
\noindent Considering that $D$ is an diagonal matrix, we will easily to get $D^{-\frac{1}{2}}$. And based on this, we get the Laplace Matrix with self-loop $L$:\\
$$
L = D - W
$$
and normalized Laplace Matrix with self-loop:\\
\begin{equation}
  L_{normal}=D^{-\frac{1}{2}}(D-W)D^{-\frac{1}{2}}
  \tag{S1}
\end{equation}
\noindent with both matrix positive semidefinite.\\

\noindent Based on this, we then do eigenvalue decomposition on the normalized Laplace Matrix with self-loop, and acquire $N$ eigenvalues and corresponding $N$ eigenvectors. Generally Speaking, we will select the first $k$ eigenvalues and the corresponding eigenvectors, which indicates that we promote the original paper galaxy map in 2-dimension to the k-dimension space. To adequately exploit the data, we choose $k = n$ for later analysis, while $k$ can be selected from $2$ to $n$ in real-world to reduce the calculation cost of huge amount of data. Then for any two papers $v_i$ and $v_j$ in topic citation network, the distance of them in k-dimensional space is $d_{ij}=\left\Vert eigvector_{v_i}-eigvector_{v_j}\right\Vert _{2}$, and we also get the distance matrix $d$ of the papers in k-dimensional space:\\
$$
d=\left(\begin{array}{cccc}
d_{11} & d_{12} & \ldots & d_{1n}\\
d_{21} & d_{22} & \ldots & d_{2n}\\
\vdots & \vdots & \ddots & \vdots\\
d_{n1} & d_{n2} & \ldots & d_{nn}
\end{array}\right)
$$

\noindent Besides, we define $MaxDistance=\max _{i,j}\left\{ d_{ij}A_{ij}\right\}$, which can be intuitively understand as the maximum distance in high-dimensional space for all edges existed in networks.\\

\noindent Based on the distance measurement in high-dimensional space, we exploit the random walk algorithm to calculate the reduction index of every node to the entire network. Whereas, before defining the reduction index of a specific node to the entire network, we need define the reduction index of single node to another (node pair) first.\\

\subsubsection*{The reduction index of one node to another (nodepair)}

\noindent For nodepair $(v_i,v_j)$, we define the reduction index of $v_i$ to $v_j$ as the sum of the weighted dijkstra path from $v_i$ to all $v_{j_{k}}$ in $v_j$'s reference list.\\
\begin{equation}
  Reduction Index_{v_i,v_j}=\underset{v_{j_{k}}}{\sum}dijkstra_{v_i,v_{j_{k}}}
  \tag{S2}
\end{equation}
\noindent Specially, for any $v_{j_{k}}$,the weighted dijkstra path from $v_i$ to $v_{j_{k}}$ is weighted sum of edges in dijkstra path when there exists path from $v_i$ to $v_{j_{k}}$, and $MaxDistance \times AverageStep$ when there exists no path from $v_i$ to $v_{j_{k}}$, where $AverageStep$ is the average of the number of steps (edges) between every nodepair that can reach each other, whether in single step or multiple steps.\\

\noindent Considering paper $v_j$, every $v_{j_{k}}$ in $v_j$'s reference list causes some influence on the paper $v_j$. In such conditions, if there exists a path from $v_i$ to $v_{j_{k}}$, which also indicates that there also exists at least one path from $v_i$ to $v_j$, and $v_i$ also somewhat causes some influence on $v_j$. And the more $v_i$ cause influence to $v_j$, the more there exists similarities between $v_i$ and $v_j$ in whether research content or idea or the method.\\

\noindent Besides, for the two situations depending on whether there exists path from $v_i$ to $v_{j_{k}}$ or not, interpretations are as follows.\\

\begin{enumerate}
	\item For the situation that there exists path from $v_i$ to $v_{j_{k}}$, such dijkstra path should quantify the distance between $v_i$ and $v_{j_{k}}$, whether they are connected directly or indirectly via paths.
	\item For the situation that there exists no path from $v_i$ to $v_{j_{k}}$, the distance calculated should be significantly larger than the distance when there exists path. Therefore, we use the $MaxDistance$ to make the value significantly large. And for the interpretation of the value of $MaxDistance$, we deem that the distance of a virtual edge (not really existing in the network) should be also significant larger than the distance of really existing edges. And for $AverageStep$, it can be interpreted that such $MaxDistance$ is for a direct virtual edge, while the virtual path from $v_i$ to $v_{j_{k}}$ may consists not only single virtual edge but multiple virtual edges that connected indirectly.
\end{enumerate}

\noindent After defining the reduction index of single node to another (node pair), we will then define the reduction index of the entire network for any specific node.\\

\subsubsection*{The reduction index of any specific node to the entire network}

\noindent For node $v$, its reduction index to the entire network $G$ is defined as the sum of its reduction index to all other nodes in the network.\\

\begin{equation}
Reduction Index_{v,G}=\underset{u \in V/{v}}{\sum}Reduction Index_{v,u}
  \tag{S3}
\end{equation}

\noindent Reduction index to the entire network helps us judge the importance of citations. The greater of the difference in the reduction index of two nodes to the network is, the less important the reference relationship between them is. Therefore, we find undirected loops and cut the unimportant nodepair according to the difference in reduction index to the Entire network while maintaining the connectivity in Directed Graph conditions. During the process, two fundamental but significant principles should be followed:\\

\begin{enumerate}
	\item Cut the nodepair with largest difference in reduction index to the entire network.
	\newline
	For this principle, we sort the nodepairs according to the difference in reduction index to the entire network, and attempt to cut them in descending order. The specific criteria to whether cut it or not is stated in pricinple 2. 
	\item Maintain the connectivity in Directed Graph conditions
	\newline For this pricinple, we do not cut the edges that represents the last reference relationship still existing in the graph, and we will skip such edges when sorting and selecting the edges to cut. In such conditions, if every article except pioneering paper reserves only one reference relationship, then the output we get after cutting the edges will be undoubtedly a tree structure.
\end{enumerate}

\noindent Following the two pricinples above, we do obtain an idea tree that reflects the idea flow of the topic citation network, which satisfies two properties:\\

\begin{enumerate}
	\item The pioneering work is the only root node of the idea tree, while the whole structure is rooted on the pioneering node.
	\item Starting from the root node of the idea tree, by doing inverse traversing operation of citation relationship, every node, or every paper in the research topic, can be reached during the traversing.\\
\end{enumerate}

\subsection*{S2.2 The calculation of knowledge entropy}
\noindent \\
\noindent \noindent Based on the idea tree, we can start from the structure of the tree and utilize structural information to measure the knowledge quality of academic articles. Specifically, the reason why we choose entropy to measure the quality of knowledge is that the entropy can measure the influence on uncertainty of the paper, which is compared with the situation that the paper does not exist. The uncertainty exists between two situations: With the paper involved, the structure of the idea tree is determined slightly; And without the paper involved, some structure of the idea tree is still unknown. Therefore, the larger a paper influences the research topic, the structure of the idea tree is more certain, and the larger its Knowledge Entropy is. \\

\subsubsection*{Subtree entropy}
\noindent For an academic paper $a$, the Subtree Entropy of $a$ is defined as follows:\\
\begin{equation}
H\left(a\right)=-\frac{g_{a}}{2m}\log\left(\frac{V_{a}}{V_{a^-}}\right)
  \tag{S4}
\end{equation}

\noindent where $g_{a}$ represents the number of the cut edges between the nodes in the subtree rooted on $a$ in the idea tree and the nodes out of the subtree. $V_{a}$ represents the number of the nodes in subtree rooted on $a$ while $V_{a^-}$ represents the number of the nodes in subtree rooted on $a$'s parent node. Generally speaking, it satisifies the relationship that $V_{a^-} \geq V_{a}+1$.\\

\subsubsection*{Mutual knowledge entropy and conditional knowledge entropy}
\noindent With the definition of subtree entropy above, the definition of mutual knowledge entropy is also given as follows:\\
\begin{equation}
I\left(a,b\right)=-\frac{g_{ab}}{4m}\log\left(\frac{V_{a}V_{b}}{V_{{ab}^-}^{2}}\right)
  \tag{S5}
\end{equation}
\noindent where $g_{ab}$ represents the number of the cut edges between the nodes in the subtree rooted on $a$ and the subtree rooted on $b$ in the idea tree acquired above and the nodes out of the two subtree. $V_{a}$ represents the number of the nodes in subtree rooted on $a$, $V_{b}$ represents the number of the nodes in subtree rooted on $b$, while $V_{{ab}^-}$ represents the number of the nodes in subtree rooted on $a$ and $b$'s parent node, which indicates that $a$ and $b$ should have the same parent node, or the two nodes should locate in similar positions in the idea tree.\\

\noindent Considering the definition form and the character of mutual knowledge entropy, it satisfies:\\
$$
I\left(a,b\right)=I\left(b,a\right)
$$
which reflects the symmetry of mutual knowledge entropy.\\
$$
I\left(a,b\right)=H\left(a\right)
$$
\noindent which reflects the self-symmetry of mutual knowledge entropy.\\

\noindent With the mutual knowledge entropy defined above, the conditional knowledge entropy is further defined:\\
$$
H\left(a\mid b\right)=H\left(a\right)-I\left(a,b\right)=\frac{g_{ab}}{4m}\log\frac{V_{a}V_{b}}{V_{{ab}^-}^{2}}-\frac{g_{a}}{2m}\log\frac{V_{a}}{V_{a^-}}
$$
\noindent And\\
$$
H\left(a,b\right)=H\left(a\right)+H\left(b\right)-I\left(a,b\right)=-\frac{g_{a}-g_{ab}}{2m}\log\frac{V_{a}}{V_{a^-}}-\frac{g_{b}-g_{ab}}{2m}\log\frac{V_{b}}{V_{a^-}}
$$
\subsubsection*{Knowledge entropy}
\noindent Based on the subtree entropy and mutual knowledge entropy above, the definition of knowledge entropy is given as follows: at timestamp $t$, the topic citation network is $G^t=\left(V^t,E^t\right)$. The idea tree extracted from the network is $IdeaTree\left(G^t\right)$. For any paper $v$ belong to $IdeaTree\left(G^t\right)$ except the seminal work, we define its knowledge entropy ${KE}^t\left(v\right)$ as:
\begin{equation}
  {KE}^t\left(v\right)=\left|H^t\left(v\right)-\sum_{v_i\in C^t\left(v\right)}{H^t\left(v_i\right)}+\sum_{v_i,v_j\in C^t\left(v\right),i\neq j}{I^t\left(v_i,v_j\right)}\right|  
  \tag{S6}
\end{equation}
\noindent where $H^t\left(v\right)$ represents the subtree entropy of the subtree led by node $v$ at $t$, $C^t\left(v\right)$ represents the children node of $v$ in $IdeaTree\left(G^t\right)$ at $t$, and $I^t\left(v_i,v_j\right)$ represents the mutual knowledge entropy of the subtree led by $v_i$ and $v_j$ at $t$.\\
\noindent As for the leading article of the topic, since it has no parent node, its subtree entropy cannot be calculated directly. However, considering the numerical difference between subtree entropy and knowledge entropy, the influnence of subtree entropy on knowledge entropy can be ignored. Therefore, for the leading article $v_s$ of any scientific topic, its knowledge entropy is given as follows.
\begin{equation}
  {KE}^t\left(v_s\right)=\left|\sum_{v_i\in C^t\left(v_s\right)}{H^t\left(v_i\right)}-\sum_{v_i,v_j\in C^t\left(v_s\right),i\neq j}{I^t\left(v_i,v_j\right)}\right|  
  \tag{S7}
\end{equation}
\noindent where $C^t\left(v_s\right)$ represents the children node of leading article in $IdeaTree\left(G^t\right)$ at $t$.\\

\setcounter{figure}{0}

%定义编号格式，在数字序号前加字符“A"

\renewcommand{\thefigure}{S\arabic{figure}}

\section*{S3 EXAMPLES OF SIX EVOLUTION PATTERNS}
\subsection*{S3.1 Pattern 1: The visible depth of the topic pioneered by summative work is hard to exceed one}

\subsubsection*{Principal component analysis: a review and recent developments}
    \begin{figure}[H]
    	\centering
    	\includegraphics[height=3.2cm]{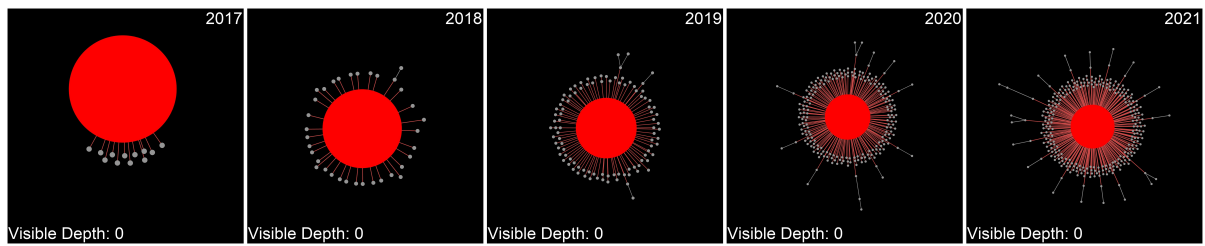}
    	\captionsetup{labelfont=bf}
    	\caption{The evolution of idea tree structure led by `Principal component analysis: a review and recent developments'}
    	\label{fig_1}
    \end{figure}

\noindent The pioneering work of the topic is `Principal component analysis: a review and recent developments'. It was published in 2016, and it already has attracted 1203 citations until 2021. Pioneering work tends to summarize existing knowledge, so it is not very inspiring for child nodes and cannot provide new research ideas. By observing the evolution of the idea tree over time, we find that even though the size of the topic continues to increase, it never breeds new high-impact nodes within it, which stagnates its visible depth at 0.\\

\subsubsection*{Multilevel Analysis : Techniques and Applications, Third Edition}
    \begin{figure}[H]
    	\centering
    	\includegraphics[height=3.2cm]{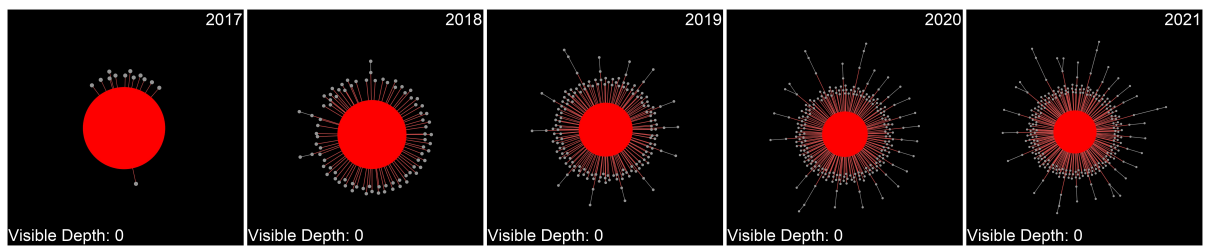}
    	\captionsetup{labelfont=bf}
    	\caption{The evolution of idea tree structure led by `Multilevel Analysis : Techniques and Applications, Third Edition'}
    	\label{fig_2}
    \end{figure}

\noindent The pioneering work of the topic is `Multilevel Analysis : Techniques and Applications, Third Edition'. It was published in 2017, and it already has attracted 2321 citations until 2021. Pioneering work tends to summarize existing knowledge, so it is not very inspiring for child nodes and cannot provide new research ideas. By observing the evolution of the idea tree over time, we find that even though the size of the topic continues to increase, it never breeds new high-impact nodes within it, which stagnates its visible depth at 0.\\

\subsubsection*{An Introduction to Medical Statistics}
    \begin{figure}[H]
    	\centering
    	\includegraphics[height=3.2cm]{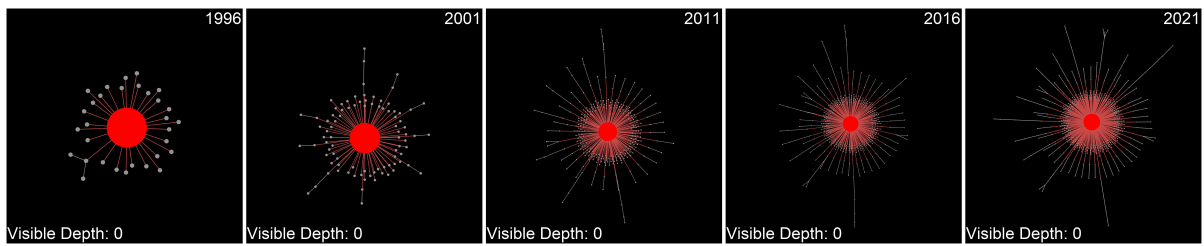}
    	\captionsetup{labelfont=bf}
    	\caption{The evolution of idea tree structure led by `An Introduction to Medical Statistics'}
    	\label{fig_3}
    \end{figure}

\noindent The pioneering work of the topic is `An Introduction to Medical Statistics'. It was published in 1987, and it already has attracted 2128 citations until 2021. Pioneering work tends to summarize existing knowledge, so it is not very inspiring for child nodes and cannot provide new research ideas. By observing the evolution of the idea tree over time, we find that even though the size of the topic continues to increase, it never breeds new high-impact nodes within it, which stagnates its visible depth at 0.\\

\subsubsection*{TensorFlow: a system for large-scale machine learning}
    \begin{figure}[H]
    	\centering
    	\includegraphics[height=3.2cm]{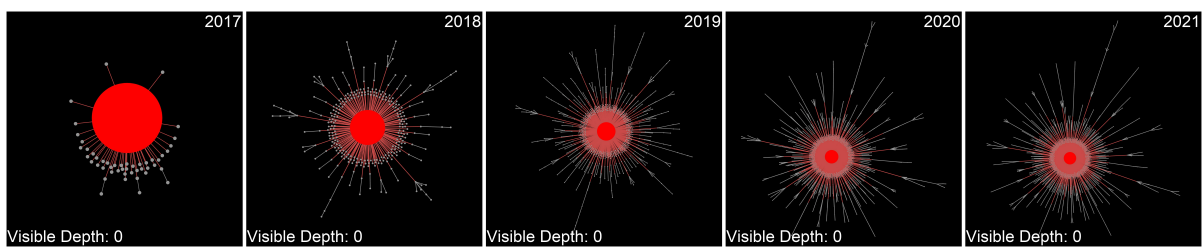}
    	\captionsetup{labelfont=bf}
    	\caption{The evolution of idea tree structure led by `TensorFlow: a system for large-scale machine learning'}
    	\label{fig_4}
    \end{figure}

\noindent The pioneering work of the topic is `TensorFlow: a system for large-scale machine learning'. It was published in 2016, and it already has attracted 5199 citations until 2021. Pioneering work tends to summarize existing knowledge, so it is not very inspiring for child nodes and cannot provide new research ideas. By observing the evolution of the idea tree over time, we find that even though the size of the topic continues to increase, it never breeds new high-impact nodes within it, which stagnates its visible depth at 0.\\

\subsection*{S3.2 Pattern 2: The increase in visible depth needs to be driven by non-trivial child nodes}

\subsubsection*{Recent Arctic amplification and extreme mid-latitude weather}
    \begin{figure}[H]
    	\centering
    	\includegraphics[height=3.2cm]{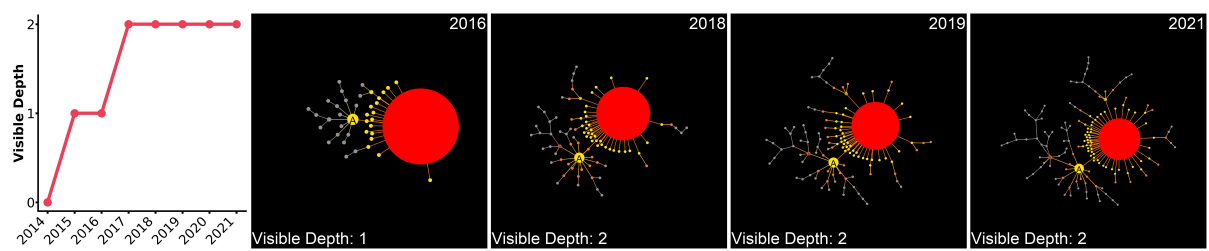}
    	\captionsetup{labelfont=bf}
    	\caption{The evolution of idea tree structure led by `Recent Arctic amplification and extreme mid-latitude weather'}
    	\label{fig_1}
    \end{figure}

\setcounter{table}{0}

%定义编号格式，在数字序号前加字符“A"

\renewcommand{\thetable}{S3-\arabic{table}}

\begin{table}[H]
    \centering
\begin{tabular}{p{2.5cm}p{9cm}p{1.5cm}p{1cm}}
\toprule
Label &                                                       Title &  KE & Year \\
\midrule
    leading article & Recent Arctic amplification and extreme mid-latitude weather & 4349.1294 &  2014\\
    A &                                                         Robust Arctic sea-ice influence on the frequent Eurasian cold winters in past decades &           201.2577 &           2014 \\
    B &                                                    Two Distinct Influences of Arctic Warming on Cold Winters over North America and East Asia &            43.0403 &           2015 \\
    C &  The impact of Arctic warming on the midlatitude jet-stream: Can it? Has it? Will it?: Impact of Arctic warming on the midlatitude jet-stream &            29.9296 &           2015 \\
\bottomrule
\end{tabular}
\captionsetup{labelfont=bf}
        \caption{Child article details}
		\label{tab_s3_1} 
\end{table}

\noindent The pioneering work of the topic is `Recent Arctic amplification and extreme mid-latitude weather'. It was published in 2014, and it already has attracted 1036 citations until 2021. Child node A attracts much external attention, making the subtree led by it ﬂourish and causing new high knowledge entropy node B to be born under its subtree, which increases the visible depth to two. This shows that paper A creates valuable knowledge and drives the topic forward. \\

\subsubsection*{Processing seismic ambient noise data to obtain reliable broad-band surface wave dispersion measurements}
    \begin{figure}[H]
    	\centering
    	\includegraphics[height=3.2cm]{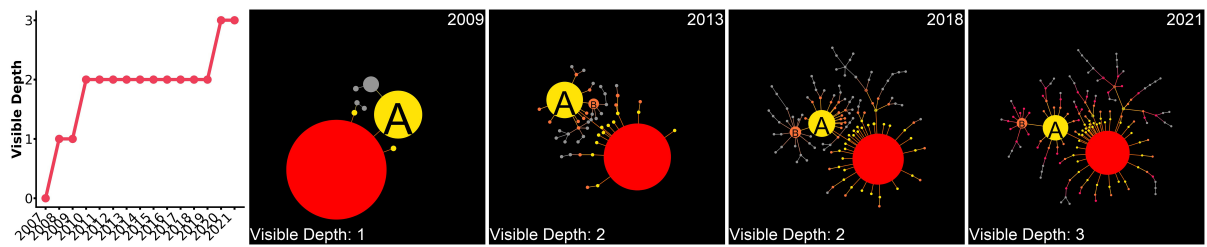}
    	\captionsetup{labelfont=bf}
    	\caption{The evolution of idea tree structure led by `Processing seismic ambient noise data to obtain reliable broad-band surface wave dispersion measurements'}
    	\label{fig_1}
    \end{figure}

\begin{table}[H]
    \centering
\begin{tabular}{p{2.5cm}p{9cm}p{1.5cm}p{1cm}}
\toprule
Label &                                                       Title &  KE & Year \\
\midrule
    leading article & Processing seismic ambient noise data to obtain reliable broad-band surface wave dispersion measurements & 4116.5964 &  2007\\
    A &                                                                        Ambient noise Rayleigh wave tomography of New Zealand &           599.2055 &           2007 \\
    B &  Surface wave tomography of the western United States from ambient seismic noise: Rayleigh and Love wave phase velocity maps &           214.5791 &           2008 \\
    C &                                                          Earthquake ground motion prediction using the ambient seismic field &            19.0616 &           2008 \\
    D &                                    Testing Community Velocity Models for Southern California Using the Ambient Seismic Field &            16.4739 &           2008 \\
    E &               Using instantaneous phase coherence for signal extraction from ambient noise data at a local to a global scale &            14.6178 &           2011 \\
    F &                                               Tutorial on seismic interferometry: Part 1 — Basic principles and applications &            10.1520 &           2010 \\
\bottomrule
\end{tabular}
\captionsetup{labelfont=bf}
        \caption{Child article details}
		\label{tab_s3_2} 
\end{table}

\noindent The pioneering work of the topic is `Processing seismic ambient noise data to obtain reliable broad-band surface wave dispersion measurements'. It was published in 2007, and it already has attracted 1307 citations until 2021. Child node A attracts much external attention, making the subtree led by it ﬂourish and causing new high knowledge entropy node B to be born under its subtree, which increases the visible depth to two. This shows that paper A creates valuable knowledge and drives the topic forward. \\

\subsubsection*{Overcoming catastrophic forgetting in neural networks}
    \begin{figure}[H]
    	\centering
    	\includegraphics[height=3.2cm]{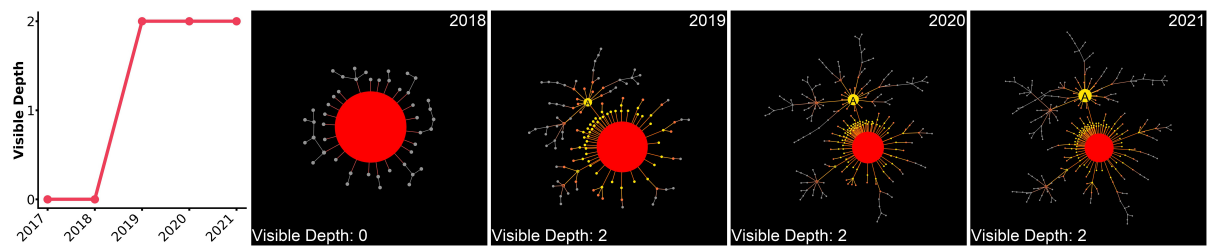}
    	\captionsetup{labelfont=bf}
    	\caption{The evolution of idea tree structure led by `Overcoming catastrophic forgetting in neural networks'}
    	\label{fig_1}
    \end{figure}

\begin{table}[H]
    \centering
\begin{tabular}{p{2.5cm}p{9cm}p{1.5cm}p{1cm}}
\toprule
Label &                                                       Title &  KE & Year \\
\midrule
    leading article & Overcoming catastrophic forgetting in neural networks & 6147.1529 &  2017\\
    A &   iCaRL: Incremental Classifier and Representation Learning &           466.7194 &           2017 \\
    B &            Continual learning through synaptic intelligence &            47.7389 &           2017 \\
    C &                             Encoder Based Lifelong Learning &            59.0671 &           2017 \\
    D &                                 Learning without Forgetting &            38.0666 &           2018 \\
    E &  Continual Lifelong Learning with Neural Networks: A Review &            23.7898 &           2019 \\
\bottomrule
\end{tabular}
\captionsetup{labelfont=bf}
        \caption{Child article details}
		\label{tab_s3_3} 
\end{table}

\noindent The pioneering work of the topic is `Overcoming catastrophic forgetting in neural networks'. It was published in 2017, and it already has attracted 1439 citations until 2021. Child node A attracts much external attention, making the subtree led by it ﬂourish and causing two new high knowledge entropy node C, E to be born under its subtree, which increases the visible depth to two. This shows that paper A creates valuable knowledge and drives the topic forward. \\

\subsubsection*{Globally and locally consistent image completion}
    \begin{figure}[H]
    	\centering
    	\includegraphics[height=3.2cm]{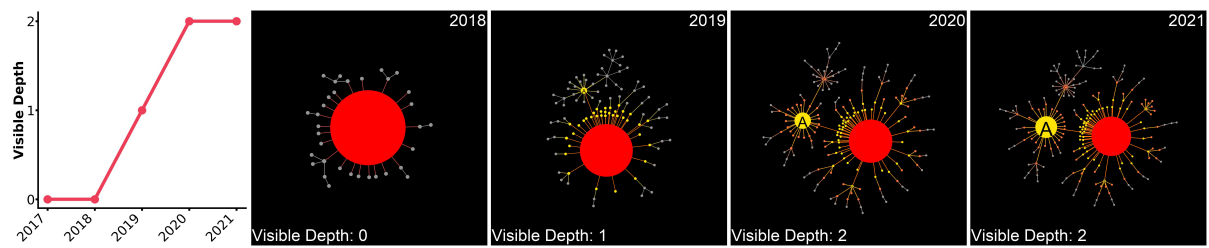}
    	\captionsetup{labelfont=bf}
    	\caption{The evolution of idea tree structure led by `Globally and locally consistent image completion'}
    	\label{fig_1}
    \end{figure}
    
\begin{table}[H]
    \centering
\begin{tabular}{p{2.5cm}p{9cm}p{1.5cm}p{1cm}}
\toprule
Label &                                                       Title &  KE & Year \\
\midrule
    leading article & Globally and locally consistent image completion & 2992.0601 &  2017\\
    A &                            Generative Image Inpainting with Contextual Attention &           567.8963 &           2018 \\
    B &                            Generative Image Inpainting with Contextual Attention (arXiv) &            16.9506 &           2018 \\
    C &                                Free-Form Image Inpainting With Gated Convolution &            64.8233 &           2018 \\
    D &  High-Resolution Image Synthesis and Semantic Manipulation with Conditional GANs &            10.2987 &           2018 \\
    E &                  Image Inpainting for Irregular Holes Using Partial Convolutions &            11.9479 &           2018 \\
\bottomrule
\end{tabular}
\captionsetup{labelfont=bf}
        \caption{Child article details}
		\label{tab_s3_4} 
\end{table}

\noindent The pioneering work of the topic is `Globally and locally consistent image completion'. It was published in 2017, and it already has attracted 1046 citations until 2021. Child node A attracts much external attention, making the subtree led by it ﬂourish and causing new high knowledge entropy node C to be born under its subtree, which increases the visible depth to two. This shows that paper A creates valuable knowledge and drives the topic forward. \\

\subsection*{Pattern 3: The continuous increase of the visible depth of the topic needs to be stimulated by the influence relay of multiple high knowledge entropy nodes}

\subsubsection*{Improved techniques for training GANs}
    \begin{figure}[H]
    	\centering
    	\includegraphics[height=3.2cm]{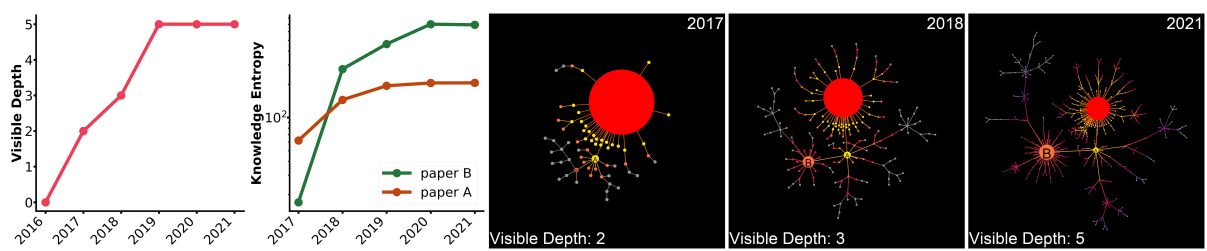}
    	\captionsetup{labelfont=bf}
    	\caption{The evolution of idea tree structure led by `Improved techniques for training GANs'}
    	\label{fig_1}
    \end{figure}
    
\begin{table}[H]
    \centering
\begin{tabular}{p{2.5cm}p{9cm}p{1.5cm}p{1cm}}
\toprule
Label &                                                       Title &  KE & Year \\
\midrule
    leading article & Improved techniques for training GANs & 5339.2978 &  2016\\
    A &                                                    Energy-based Generative Adversarial Networks &           206.1023 &           2017 \\
    B &                                Image-to-Image Translation with Conditional Adversarial Networks &           692.8008 &           2017 \\
    C &                                        Calibrating Energy-based Generative Adversarial Networks &            13.0860 &           2017 \\
    D &                                                           Improved Training of Wasserstein GANs (arXiv) &            22.7750 &           2017 \\
    E &                    Learning from Simulated and Unsupervised Images through Adversarial Training &            23.9048 &           2017 \\
    F &                                                           Improved Training of Wasserstein GANs &            34.6206 &           2017 \\
    G &                 Unpaired Image-to-Image Translation Using Cycle-Consistent Adversarial Networks &            41.4849 &           2017 \\
    H &  StackGAN: Text to Photo-Realistic Image Synthesis with Stacked Generative Adversarial Networks &            17.4041 &           2017 \\
    I &                Learning to Discover Cross-Domain Relations with Generative Adversarial Networks &            25.9975 &           2017 \\
\bottomrule
\end{tabular}
\captionsetup{labelfont=bf}
        \caption{Child article details}
		\label{tab_s3_5} 
\end{table}

\noindent The pioneering work of the topic is `Improved techniques for training GANs'. It was published in 2016, and it already has attracted 2247 citations until 2021. The knowledge entropy of child node A is the ﬁrst to highlight, but after 2018, its knowledge entropy almost stopped increasing. At this time, the knowledge entropy of the child node B, which was directly inspired by paper A, has begun to emerge and exceed paper A. Paper B took over the task of motivating the visible depth increase. The topic achieves multiple inheritances of ideas and ensures that it can continue to attract attention. Several child nodes with high knowledge entropy were birthed under the subtree led by paper B, and made the visible depth of the topic continuously increased to 5. \\

\subsubsection*{Matching networks for one shot learning}
    \begin{figure}[H]
    	\centering
    	\includegraphics[height=3.2cm]{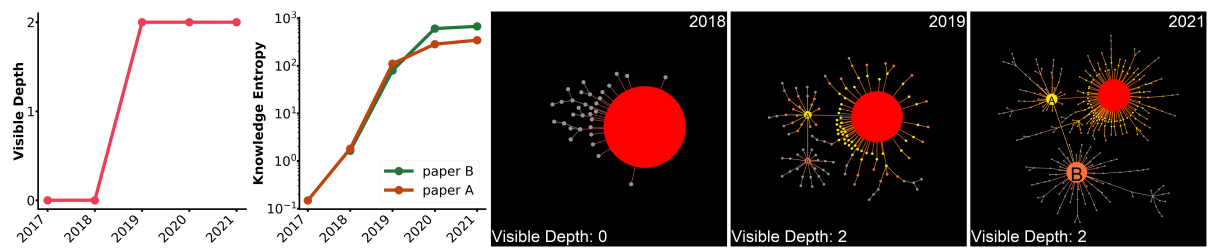}
    	\captionsetup{labelfont=bf}
    	\caption{The evolution of idea tree structure led by `Matching networks for one shot learning'}
    	\label{fig_1}
    \end{figure}

\begin{table}[H]
    \centering
\begin{tabular}{p{2.5cm}p{9cm}p{1.5cm}p{1cm}}
\toprule
Label &                                                       Title &  KE & Year \\
\midrule
    leading article & Matching networks for one shot learning & 6619.5201 &  2016\\
    A &                                                Meta Networks &           345.9163 &           2017 \\
    B &  Learning to Compare: Relation Network for Few-Shot Learning &           677.4604 &           2018 \\
\bottomrule
\end{tabular}
\captionsetup{labelfont=bf}
        \caption{Child article details}
		\label{tab_s3_6} 
\end{table}

\noindent The pioneering work of the topic is `Matching networks for one shot learning'. It was published in 2016, and it already has attracted 1515 citations until 2021. The knowledge entropy of child node A is the ﬁrst to highlight, but after 2019, its knowledge entropy almost stopped increasing. At this time, the knowledge entropy of the child node B, which was directly inspired by paper A, has begun to emerge and exceed paper A. Paper B took over the task of motivating the visible depth increase. The topic achieves multiple inheritances of ideas from the leading article and ensures that it can continue to attract attention. This makes the visible depth of the topic continuously increased to 2. \\

\subsubsection*{A Meta-Analysis of Global Urban Land Expansion}
    \begin{figure}[H]
    	\centering
    	\includegraphics[height=3.2cm]{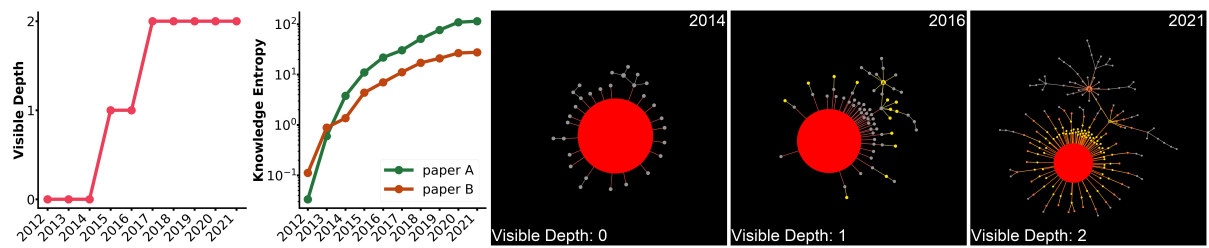}
    	\captionsetup{labelfont=bf}
    	\caption{The evolution of idea tree structure led by `A Meta-Analysis of Global Urban Land Expansion'}
    	\label{fig_1}
    \end{figure}

\begin{table}[H]
    \centering
\begin{tabular}{p{2.5cm}p{9cm}p{1.5cm}p{1cm}}
\toprule
Label &                                                       Title &  KE & Year \\
\midrule
    leading article & Matching networks for one shot learning & 2673.4137 &  2011\\
    A &  Global forecasts of urban expansion to 2030 and direct impacts on biodiversity and carbon pools &           114.8178 &           2012 \\
    B &                                                    Urban land teleconnections and sustainability &            27.6795 &           2012 \\
\bottomrule
\end{tabular}
\captionsetup{labelfont=bf}
        \caption{Child article details}
		\label{tab_s3_7} 
\end{table}

\noindent The pioneering work of the topic is `A Meta-Analysis of Global Urban Land Expansion'. It was published in 2011, and it already has attracted 1118 citations until 2021. The knowledge entropy of child node B is the ﬁrst to highlight, but after 2013, The growth of its knowledge entropy started to slow down. At this time, the knowledge entropy of the child node A, which was directly inspired by paper B, has begun to emerge and exceed paper B. Paper A took over the task of motivating the visible depth increase. The topic achieves multiple inheritances of ideas from the leading article and ensures that it can continue to attract attention. This mades the visible depth of the topic continuously increased to 2. \\

\subsubsection*{Cleavage of GSDMD by inflammatory caspases determines pyroptotic cell death}
    \begin{figure}[H]
    	\centering
    	\includegraphics[height=3.2cm]{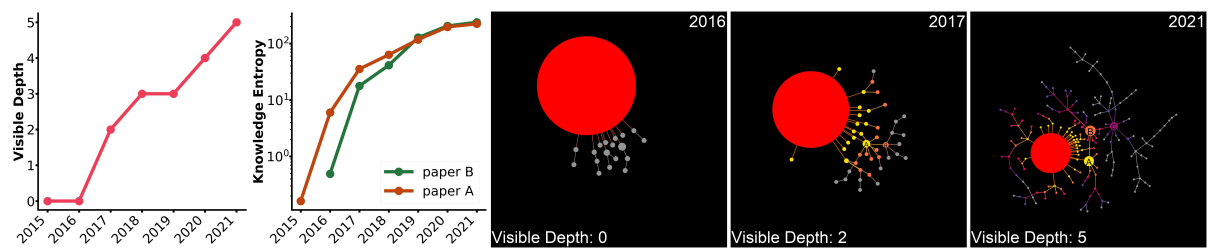}
    	\captionsetup{labelfont=bf}
    	\caption{The evolution of idea tree structure led by `Cleavage of GSDMD by inflammatory caspases determines pyroptotic cell death'}
    	\label{fig_1}
    \end{figure}

\begin{table}[H]
    \centering
\begin{tabular}{p{2.5cm}p{9cm}p{1.5cm}p{1cm}}
\toprule
Label &                                                       Title &  KE & Year \\
\midrule
    leading article & Cleavage of GSDMD by inflammatory caspases determines pyroptotic cell death & 2864.9248 &  2015\\
    A &                                                                                  Gasdermin D is an executor of pyroptosis and required for interleukin-1$\beta$ secretion &           223.7697 &           2015 \\
    B &                                                                                      Inflammasome-activated gasdermin D causes pyroptosis by forming membrane pores &           238.3985 &           2016 \\
    C &                                                                                      Chemotherapy drugs induce pyroptosis through caspase-3 cleavage of a gasdermin &           167.2502 &           2017 \\
    D &  A Genome-wide CRISPR (Clustered Regularly Interspaced Short Palindromic Repeats) Screen Identifies NEK7 as an Essential Component of NLRP3 Inflammasome Activation &            14.3097 &           2016 \\
    E &                                      Pyroptosis is driven by non-selective gasdermin-D pore and its morphology is different from MLKL channel-mediated necroptosis. &            11.3740 &           2016 \\
    F &                                                                                                                                 Gasdermins: Effectors of Pyroptosis &            18.3531 &           2017 \\
\bottomrule
\end{tabular}
\captionsetup{labelfont=bf}
        \caption{Child article details}
		\label{tab_s3_8} 
\end{table}

\noindent The pioneering work of the topic is `Cleavage of GSDMD by inflammatory caspases determines pyroptotic cell death'. It was published in 2015, and it already has attracted 1583 citations until 2021. The knowledge entropy of child node A is the ﬁrst to highlight, but after 2019, the knowledge entropy of the child node B, which was directly inspired by paper A, has begun to exceed paper A. Paper B took over the task of motivating the visible depth increase. The topic achieves multiple inheritances of ideas and ensures that it can continue to attract attention. Several child nodes with high knowledge entropy were birthed under the subtree led by paper B, and made the visible depth of the topic continuously increased to 5. \\

\subsection*{Pattern 4: The presence of overpowered child nodes can ruin the increase in the visible depth of the topic}

\subsubsection*{The Kinetics Human Action Video Dataset}
    \begin{figure}[H]
    	\centering
    	\includegraphics[height=3.2cm]{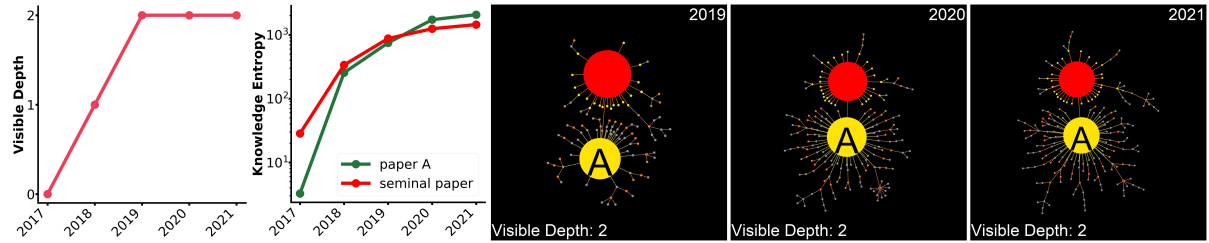}
    	\captionsetup{labelfont=bf}
    	\caption{The evolution of idea tree structure led by `The Kinetics Human Action Video Dataset'}
    	\label{fig_1}
    \end{figure}

\begin{table}[H]
    \centering
\begin{tabular}{p{2.5cm}p{9cm}p{1.5cm}p{1cm}}
\toprule
Label &                                                       Title &  KE & Year \\
\midrule
    leading article & The Kinetics Human Action Video Dataset & 1447.5796 &  2017\\
    A &     Quo Vadis, Action Recognition? A New Model and the Kinetics Dataset &          1069.7566 &           2017 \\
    B &  Can Spatiotemporal 3D CNNs Retrace the History of 2D CNNs and ImageNet &            31.6628 &           2018 \\
\bottomrule
\end{tabular}
\captionsetup{labelfont=bf}
        \caption{Child article details}
		\label{tab_s3_9} 
\end{table}

\noindent The pioneering work of the topic is `The Kinetics Human Action Video Dataset'. It was published in 2017, and it already has attracted 1043 citations until 2021. The knowledge entropy of the child article A increases rapidly and approaches the order of magnitude of the leading article’s knowledge entropy. However, in the subtree led by article A, new high knowledge entropy nodes are difficult to appear in large numbers. This indicates that the original topic lost attention of new valuable knowledge and fell into a stagnation of development. \\

\subsubsection*{Estimating Corn Leaf Chlorophyll Concentration from Leaf and Canopy Reflectance}
    \begin{figure}[H]
    	\centering
    	\includegraphics[width=15.5cm]{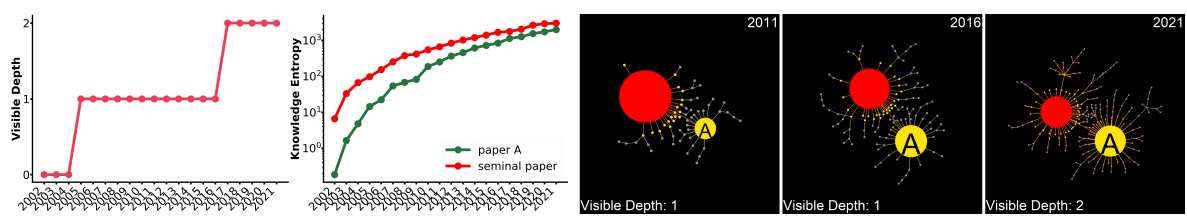}
    	\captionsetup{labelfont=bf}
    	\caption{The evolution of idea tree structure led by `Estimating Corn Leaf Chlorophyll Concentration from Leaf and Canopy Reflectance'}
    	\label{fig_1}
    \end{figure}

\begin{table}[H]
    \centering
\begin{tabular}{p{2.5cm}p{9cm}p{1.5cm}p{1cm}}
\toprule
Label &                                                       Title &  KE & Year \\
\midrule
    leading article & Estimating Corn Leaf Chlorophyll Concentration from Leaf and Canopy Reflectance & 3019.6228 &  2000\\
    A &                                     Integrated narrow-band vegetation indices for prediction of crop chlorophyll content for application to precision agriculture &          1069.0009 &           2002 \\
    B &  Hyperspectral vegetation indices and novel algorithms for predicting green LAI of crop canopies: Modeling and validation in the context of precision agriculture &            42.3769 &           2004 \\
    C &                                                        Wide Dynamic Range Vegetation Index for Remote Quantification of Biophysical Characteristics of Vegetation &            12.7616 &           2004 \\
\bottomrule
\end{tabular}
\captionsetup{labelfont=bf}
        \caption{Child article details}
		\label{tab_s3_10} 
\end{table}

\noindent The pioneering work of the topic is `Estimating Corn Leaf Chlorophyll Concentration from Leaf and Canopy Reflectance'. It was published in 2000, and it already has attracted 1281 citations until 2021. The knowledge entropy of the child article A approaches the order of magnitude of the leading article’s knowledge entropy. However, in the subtree led by article A, new high knowledge entropy nodes are difficult to appear in large numbers. This indicates that the original topic lost attention of new valuable knowledge and fell into a stagnation of development. \\

\subsubsection*{Thermal remote sensing of urban climates}
    \begin{figure}[H]
    	\centering
    	\captionsetup{labelfont=bf}
    	\includegraphics[width=15.5cm]{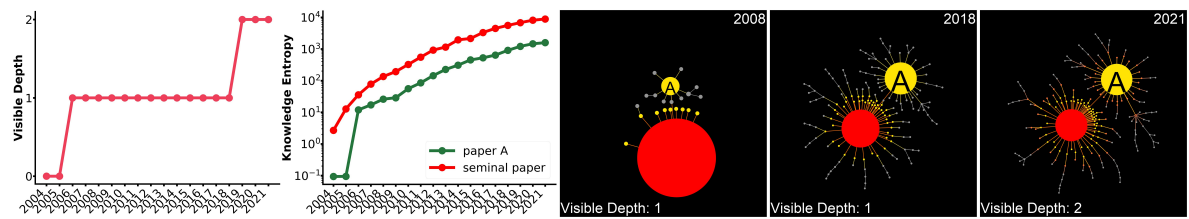}
    	\caption{The evolution of idea tree structure led by `Thermal remote sensing of urban climates'}
    	\label{fig_1}
    \end{figure}

\begin{table}[H]
    \centering
\begin{tabular}{p{2.5cm}p{9cm}p{1.5cm}p{1cm}}
\toprule
Label &                                                       Title &  KE & Year \\
\midrule
    leading article & Thermal remote sensing of urban climates & 8846.6336 &  2003\\
    A &  Estimation of land surface temperature-vegetation abundance relationship for urban heat island studies &          1063.6782 &           2004 \\
    B &                                                  Surface Urban Heat Island Across 419 Global Big Cities &            29.1588 &           2012 \\
\bottomrule
\end{tabular}
\captionsetup{labelfont=bf}
        \caption{Child article details}
		\label{tab_s3_11} 
\end{table}

\noindent The pioneering work of the topic is `Thermal remote sensing of urban climates'. It was published in 2003, and it already has attracted 1539 citations until 2021. The knowledge entropy of the child article A increases rapidly and approaches the order of magnitude of the leading article’s knowledge entropy. However, in the subtree led by article A, new high knowledge entropy nodes are difficult to appear in large numbers. This indicates that the original topic lost attention of new valuable knowledge and fell into a stagnation of development. \\

\subsubsection*{Weyl Semimetal Phase in Noncentrosymmetric Transition-Metal Monophosphides}
    \begin{figure}[H]
    	\centering
    	\includegraphics[width=15.5cm]{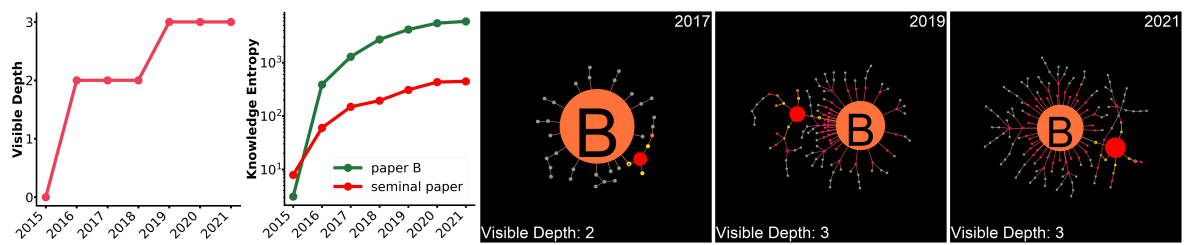}
    	\captionsetup{labelfont=bf}
    	\caption{The evolution of idea tree structure led by `Weyl Semimetal Phase in Noncentrosymmetric Transition-Metal Monophosphides'}
    	\label{fig_1}
    \end{figure}

\begin{table}[H]
    \centering
\begin{tabular}{p{2.5cm}p{9cm}p{1.5cm}p{1cm}}
\toprule
Label &                                                       Title &  KE & Year \\
\midrule
    leading article & Weyl Semimetal Phase in Noncentrosymmetric Transition-Metal Monophosphides & 446.7779 &  2015\\
    A &                                                                       Experimental observation of Weyl points &            21.0205 &           2015 \\
    B &                                                                 Experimental discovery of Weyl semimetal TaAs &          1085.0453 &           2015 \\
    C &      Extremely large magnetoresistance and ultrahigh mobility in the topological Weyl semimetal candidate NbP &             9.8317 &           2015 \\
    D &                                                                             Observation of Weyl nodes in TaAs &             7.5791 &           2015 \\
    E &  Line-Node Dirac Semimetal and Topological Insulating Phase in Noncentrosymmetric Pnictides CaAgX (X = P, As) &            13.3865 &           2016 \\
    F &                                                                             Topological nodal line semimetals &            28.1282 &           2016 \\
    G &                                                         Weyl and Dirac semimetals in three-dimensional solids &            32.1386 &           2018 \\
    H &                                                                               Triple Point Topological Metals &            27.6119 &           2016 \\
\bottomrule
\end{tabular}
\captionsetup{labelfont=bf}
        \caption{Child article details}
		\label{tab_s3_12} 
\end{table}

\noindent The pioneering work of the topic is `Weyl Semimetal Phase in Noncentrosymmetric Transition-Metal Monophosphides'. It was published in 2015, and it already has attracted 1105 citations until 2021. The knowledge entropy of the child article B increases rapidly and exceeds the order of magnitude of the leading article’s knowledge entropy. However, in the subtree led by article B, new high knowledge entropy nodes are difficult to appear in large numbers. This indicates that the original topic lost attention of new valuable knowledge and fell into a stagnation of development. \\

\subsection*{Pattern 5: Stronger branches within the topic inhibit the increase in visible depth of weaker branches}

\subsubsection*{Non-ideal interactions in calcic amphiboles and their bearing on amphibole-plagioclase thermometry}
    \begin{figure}[H]
    	\centering
    	\includegraphics[width=15.5cm]{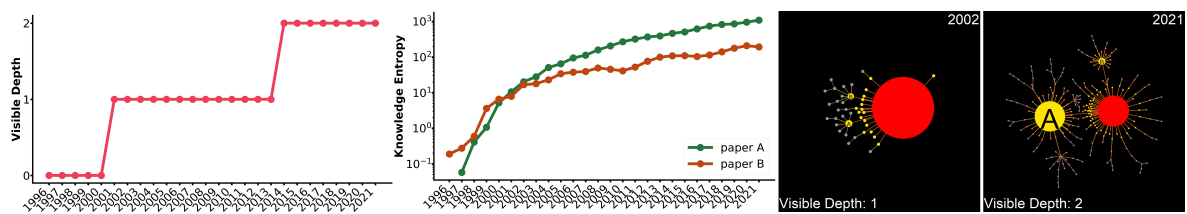}
    	\captionsetup{labelfont=bf}
    	\caption{The evolution of idea tree structure led by `Non-ideal interactions in calcic amphiboles and their bearing on amphibole-plagioclase thermometry'}
    	\label{fig_1}
    \end{figure}

\begin{table}[H]
    \centering
\begin{tabular}{p{2.5cm}p{9cm}p{1.5cm}p{1cm}}
\toprule
Label &                                                       Title &  KE & Year \\
\midrule
    leading article & Non-ideal interactions in calcic amphiboles and their bearing on amphibole-plagioclase thermometry & 2747.0623 &  1994\\
    A &  Nomenclature of amphiboles; Report of the Subcommittee on Amphiboles of the International Mineralogical Association, Commission on New Minerals and Mineral Names &          1044.7652 &           1997 \\
    B &                                                                                              The effects of temperature and f O2 on the Al-in-hornblende barometer &           192.8548 &           1995 \\
    C &                                        Experimental phase-equilibrium study of Al- and Ti-contents of calcic amphibole in MORB; a semiquantitative thermobarometer &            28.5731 &           1998 \\
    D &                                                                                    Temperature-induced Al-zoning in hornblendes of the Fish Canyon magma, Colorado &            10.9639 &           2002 \\
\bottomrule
\end{tabular}
\captionsetup{labelfont=bf}
        \caption{Child article details}
		\label{tab_s3_13} 
\end{table}

\noindent The pioneering work of the topic is `Non-ideal interactions in calcic amphiboles and their bearing on amphibole-plagioclase thermometry'. It was published in 1994, and it already has attracted 1612 citations until 2021. In the next layer of leading work, two nodes with high knowledge entropy were born. In the early stage of the topic's development, the subtree led by paper B first became prosperous. When the knowledge entropy of paper A exceeds paper B, the subtree led by paper A began to become prosperous, and new high knowledge entropy nodes were born in it, thus increasing the topic visible depth by 1. In this process, the newly emerged branches attracted the attention of the outside world. In contrast, the subtree led by paper B began to be ignored, thus missing the golden opportunity for development, leading to stagnation of its development.\\

\subsubsection*{A dynamic global vegetation model for studies of the coupled atmosphere‐biosphere system}
    \begin{figure}[H]
    	\centering
    	\includegraphics[width=15.5cm]{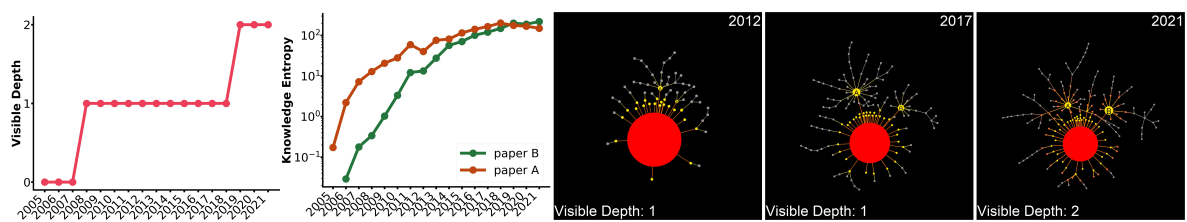}
    	\captionsetup{labelfont=bf}
    	\caption{The evolution of idea tree structure led by `A dynamic global vegetation model for studies of the coupled atmosphere‐biosphere system'}
    	\label{fig_1}
    \end{figure}

\begin{table}[H]
    \centering
\begin{tabular}{p{2.5cm}p{9cm}p{1.5cm}p{1cm}}
\toprule
Label &                                                       Title &  KE & Year \\
\midrule
    leading article & A dynamic global vegetation model for studies of the coupled atmosphere‐biosphere system & 74525.8388&  2005\\
    A &                                                   Europe-wide reduction in primary productivity caused by the heat and drought in 2003 &           150.1318 &           2005 \\
    B &                                                   Climate-carbon cycle feedback analysis: Results from the C4MIP model intercomparison &           220.0082 &           2006 \\
    C &  The LMDZ4 general circulation model: climate performance and sensitivity to parametrized physics with emphasis on tropical convection &            12.8155 &           2006 \\
    D &                                          Key features of the IPSL ocean atmosphere model and its sensitivity to atmospheric resolution &            12.4421 &           2010 \\
\bottomrule
\end{tabular}
\captionsetup{labelfont=bf}
        \caption{Child article details}
		\label{tab_s3_14} 
\end{table}

\noindent The pioneering work of the topic is `A dynamic global vegetation model for studies of the coupled atmosphere‐biosphere system'. It was published in 2005, and it already has attracted 1467 citations until 2021. In the next layer of leading work, two nodes with high knowledge entropy were born. In the early stage of the topic's development, the subtree led by paper A first became prosperous. When the knowledge entropy of paper B exceeds paper A, the subtree led by paper B began to become prosperous. In contrast, the subtree led by paper A began to be ignored, thus missing the golden opportunity for development, leading to stagnation of its development.\\

\subsubsection*{Robust Responses of the Hydrological Cycle to Global Warming}
    \begin{figure}[H]
    	\centering
    	\includegraphics[width=15.5cm]{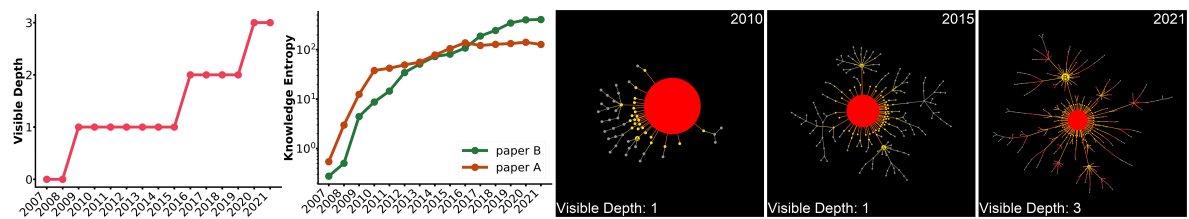}
    	\captionsetup{labelfont=bf}
    	\caption{The evolution of idea tree structure led by `Robust Responses of the Hydrological Cycle to Global Warming'}
    	\label{fig_1}
    \end{figure}

\begin{table}[H]
    \centering
\begin{tabular}{p{2.5cm}p{9cm}p{1.5cm}p{1cm}}
\toprule
Label &                                                       Title &  KE & Year \\
\midrule
    leading article & Robust Responses of the Hydrological Cycle to Global Warming & 14518.76612 &  2006\\
    A &                                             How Much More Rain Will Global Warming Bring &           127.3424 &           2007 \\
    B &                                        Expansion of the Hadley cell under global warming &           407.5004 &           2007 \\
    C &  Mechanisms for the land/sea warming contrast exhibited by simulations of climate change &            16.4972 &           2008 \\
    D &        Controls of Global-Mean Precipitation Increases in Global Warming GCM Experiments &            24.3881 &           2008 \\
    E &                                             Changes in precipitation with climate change &            75.6538 &           2011 \\
    F &            Increased tropical Atlantic wind shear in model projections of global warming &            10.1380 &           2007 \\
    G &                  The impact of global warming on the tropical Pacific Ocean and El Niño. &            16.3916 &           2010 \\
    H &                                              Effects of increased CO2 levels on monsoons &            12.4175 &           2011 \\
\bottomrule
\end{tabular}
\captionsetup{labelfont=bf}
        \caption{Child article details}
		\label{tab_s3_15} 
\end{table}

\noindent The pioneering work of the topic is Robust Responses of the Hydrological Cycle to Global Warming'. It was published in 2006, and it already has attracted 3083 citations until 2021. In the next layer of leading work, two nodes with high knowledge entropy were born. In the early stage of the topic's development, the subtree led by paper A first became prosperous. When the knowledge entropy of paper B exceeds paper A, the subtree led by paper B began to become prosperous, and new high knowledge entropy nodes were born in it, thus increasing the topic visible depth by 1. In this process, the newly emerged branches attracted the attention of the outside world. In contrast, the branches led by paper A were ignored, thus missing the golden opportunity for development, leading to stagnation of its development.\\

\subsubsection*{Prototypical Networks for Few-shot Learning}
    \begin{figure}[H]
    	\centering
    	\includegraphics[height=3.2cm]{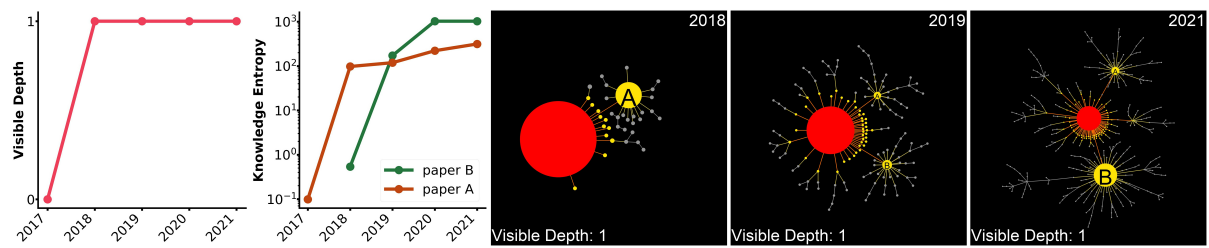}
    	\captionsetup{labelfont=bf}
    	\caption{The evolution of idea tree structure led by `Prototypical Networks for Few-shot Learning'}
    	\label{fig_1}
    \end{figure}

\begin{table}[H]
    \centering
\begin{tabular}{p{2.5cm}p{9cm}p{1.5cm}p{1cm}}
\toprule
Label &                                                       Title &  KE & Year \\
\midrule
    leading article & Prototypical Networks for Few-shot Learning & 7627.8892 &  2017\\
    A &  Model-agnostic meta-learning for fast adaptation of deep networks &           314.6073 &           2017 \\
    B &        Learning to Compare: Relation Network for Few-Shot Learning &          1033.2353 &           2018 \\
    C &                              Low-Shot Learning from Imaginary Data &            17.4369 &           2018 \\
\bottomrule
\end{tabular}
\captionsetup{labelfont=bf}
        \caption{Child article details}
		\label{tab_s3_16} 
\end{table}

\noindent The pioneering work of the topic is `Prototypical Networks for Few-shot Learning'. It was published in 2017, and it already has attracted 1964 citations until 2021. In the next layer of leading work, two nodes with high knowledge entropy were born. In the early stage of the topic's development, the subtree led by paper A first became prosperous. When the knowledge entropy of paper B exceeds paper A, the subtree led by paper B began to become prosperous. In contrast, the subtree led by paper A began to be ignored, thus missing the golden opportunity for development, leading to stagnation of its development.\\

\subsection*{Pattern 6: Visible depth near the upper bound of development requires a large number of high knowledge entropy nodes to drive}

\subsubsection*{Unsupervised Representation Learning with Deep Convolutional Generative Adversarial Networks}

\begin{figure}[H]
	\centering
	\includegraphics[height=3.2cm]{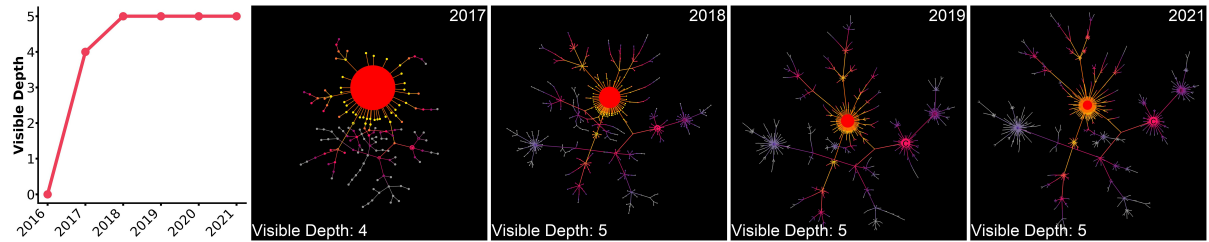}
	\captionsetup{labelfont=bf}
	\caption{The evolution of idea tree structure led by `Unsupervised Representation Learning with Deep Convolutional Generative Adversarial Networks'}
	\label{fig_1}
\end{figure}

\begin{table}[H]
    \centering
\begin{tabular}{p{2.5cm}p{9cm}p{1.8cm}p{1cm}}
\toprule
Label &                                                       Title &  KE & Year \\
\midrule
    leading article & Unsupervised Representation Learning with Deep Convolutional Generative Adversarial Networks & 21565.4562 &  2016\\
    A &  InfoGAN: interpretable representation learning by information maximizing generative adversarial nets &           141.9261 &           2016 \\
    B &                                                                 Improved techniques for training GANs &           134.6050 &           2016 \\
    C &                                                                 Improved techniques for training GANs &           978.0934 &           2016 \\
    D &                                            Conditional Image Synthesis With Auxiliary Classifier GANs &            84.4415 &           2017 \\
    E &                                      Image-to-Image Translation with Conditional Adversarial Networks &           498.4536 &           2017 \\
    F &                                  Attribute2Image: Conditional Image Generation from Visual Attributes &            53.6515 &           2016 \\
    G &                                                               Coupled generative adversarial networks &            29.1537 &           2016 \\
    H &                                   Perceptual Losses for Real-Time Style Transfer and Super-Resolution &            22.8999 &           2016 \\
    I &                                          Generative Visual Manipulation on the Natural Image Manifold &            18.2762 &           2016 \\
    J &                                                              Unrolled Generative Adversarial Networks &            44.2670 &           2016 \\
    K &                       Unpaired Image-to-Image Translation Using Cycle-Consistent Adversarial Networks &            46.5298 &           2017 \\
    L &                                                                                       Wasserstein GAN &           316.0971 &           2017 \\
    M &                                  Attribute2Image: Conditional Image Generation from Visual Attributes &            16.5627 &           2016 \\
    N &                                                                 Improved Training of Wasserstein GANs &           114.3471 &           2017 \\
    O &                                                              Deconvolution and Checkerboard Artifacts &            11.7450 &           2016 \\
    P &                 f -GAN: training generative neural samplers using variational divergence minimization &            14.1597 &           2016 \\
    Q &                                                 Semantic Image Inpainting with Deep Generative Models &            11.1524 &           2017 \\
\bottomrule
\end{tabular}
\captionsetup{labelfont=bf}
        \caption{Child article details}
		\label{tab_s3_17} 
\end{table}

\noindent The pioneering work of the topic is `Unsupervised Representation Learning with Deep Convolutional Generative Adversarial Networks'. It was published in 2016, and it already has attracted 5789 citations until 2021. A large number of high knowledge entropy nodes are bred inside the topic, which makes the visible depth of the topic reach 5. \\

\subsubsection*{Feature Pyramid Networks for Object Detection}
    \begin{figure}[H]
    	\centering
    	\includegraphics[height=3.2cm]{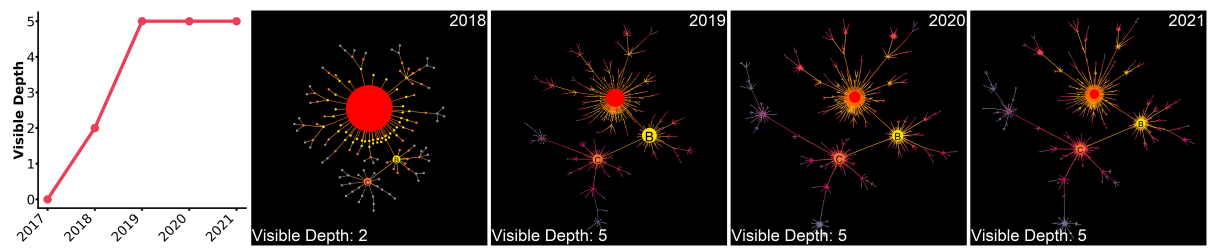}
    	\captionsetup{labelfont=bf}
    	\caption{The evolution of idea tree structure led by `Feature Pyramid Networks for Object Detection'}
    	\label{fig_1}
    \end{figure}

\begin{table}[H]
    \centering
\begin{tabular}{p{2.5cm}p{9cm}p{1.8cm}p{1cm}}
\toprule
Label &                                                       Title &  KE & Year \\
\midrule
    leading article & Feature Pyramid Networks for Object Detection & 15048.9172 &  2017\\
    A &                                              Focal Loss for Dense Object Detection (arXiv) &            59.7089 &           2017 \\
    B &                                                                         Mask R-CNN &          1072.5319 &           2017 \\
    C &                                              Focal Loss for Dense Object Detection &          1056.9098 &           2017 \\
    D &                Speed/Accuracy Trade-Offs for Modern Convolutional Object Detectors &            22.3072 &           2017 \\
    E &                                                                         Mask R-CNN (arXiv) &            19.9824 &           2017 \\
    F &                                                 YOLOv3: An Incremental Improvement &            98.4034 &           2018 \\
    G &                          Cascade R-CNN: Delving Into High Quality Object Detection &            70.0600 &           2018 \\
    H &                                   CornerNet: Detecting Objects as Paired Keypoints &           164.2707 &           2018 \\
    I &  Encoder-Decoder with Atrous Separable Convolution for Semantic Image Segmentation &            52.5990 &           2018 \\
   J &                                 Path Aggregation Network for Instance Segmentation &            12.4741 &           2018 \\
   K &                 Learning Transferable Architectures for Scalable Image Recognition &            21.7402 &           2018 \\
   L &                          Frustum PointNets for 3D Object Detection from RGB-D Data &            12.8733 &           2018 \\
\bottomrule
\end{tabular}
\captionsetup{labelfont=bf}
        \caption{Child article details}
		\label{tab_s3_18} 
\end{table}

\noindent The pioneering work of the topic is `Feature Pyramid Networks for Object Detection'. It was published in 2017, and it already has attracted 5031 citations until 2021. A large number of high knowledge entropy nodes are bred inside the topic, which makes the visible depth of the topic reach 5. \\

\subsubsection*{Inception-v4, Inception-ResNet and the Impact of Residual Connections on Learning}
    \begin{figure}[H]
    	\centering
    	\includegraphics[height=3.2cm]{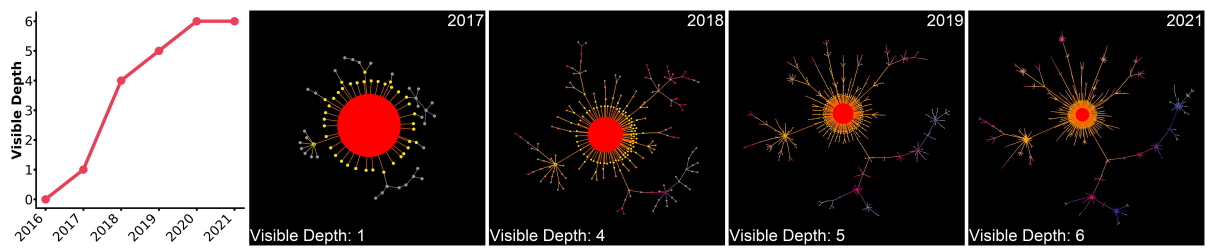}
    	\captionsetup{labelfont=bf}
    	\caption{The evolution of idea tree structure led by `Inception-v4, Inception-ResNet and the Impact of Residual Connections on Learning'}
    	\label{fig_1}
    \end{figure}
    
\begin{table}[H]
    \centering
\begin{tabular}{p{2.5cm}p{9cm}p{1.8cm}p{1cm}}
\toprule
Label &                                                       Title &  KE & Year \\
\midrule
    leading article & Inception-v4, Inception-ResNet and the Impact of Residual Connections on Learning & 11336.1737 &  2016\\
    A &                                         Identity Mappings in Deep Residual Networks &            96.9002 &           2016 \\
    B &                        Aggregated Residual Transformations for Deep Neural Networks &            20.0488 &           2016 \\
    C &                 Speed/Accuracy Trade-Offs for Modern Convolutional Object Detectors &            67.4979 &           2017 \\
    D &                                                                          Mask R-CNN (arXiv) &            25.2243 &           2017 \\
    E &  MobileNets: Efficient Convolutional Neural Networks for Mobile Vision Applications &            16.8268 &           2017 \\
    F &                                                                          Mask R-CNN &            30.6026 &           2017 \\
    G &                  Learning Transferable Architectures for Scalable Image Recognition &            36.2898 &           2018 \\
    H &                       Xception: Deep Learning with Depthwise Separable Convolutions &            23.0431 &           2017 \\
    I &                                               Focal Loss for Dense Object Detection &            13.8385 &           2017 \\
   J &                                                     Squeeze-and-Excitation Networks &            26.8498 &           2018 \\
\bottomrule
\end{tabular}
\captionsetup{labelfont=bf}
        \caption{Child article details}
		\label{tab_s3_19} 
\end{table}

\noindent The pioneering work of the topic is `Inception-v4, Inception-ResNet and the Impact of Residual Connections on Learning'. It was published in 2016, and it already has attracted 3024 citations until 2021. A large number of high knowledge entropy nodes are bred inside the topic, which makes the visible depth of the topic reach 6. \\

\subsubsection*{High Serum IgG4 Concentrations in Patients with Sclerosing Pancreatitis}
    \begin{figure}[H]
    	\centering
    	\includegraphics[width=15.5cm]{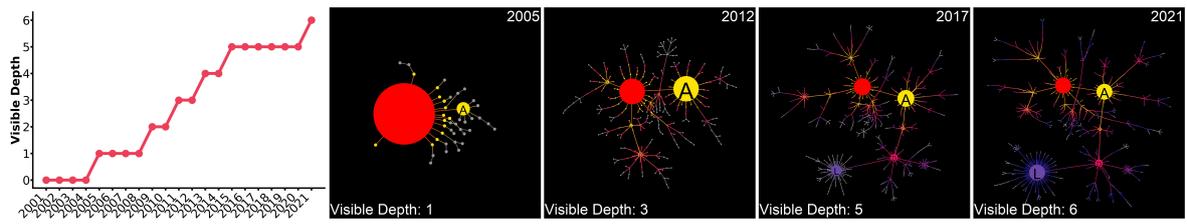}
    	\captionsetup{labelfont=bf}
    	\caption{The evolution of idea tree structure led by `High Serum IgG4 Concentrations in Patients with Sclerosing Pancreatitis'}
    	\label{fig_1}
    \end{figure}
    
\begin{table}[H]
    \centering
\begin{tabular}{p{2.5cm}p{9cm}p{1.8cm}p{1cm}}
\toprule
Label &                                                       Title &  KE & Year \\
\midrule
    leading article & High Serum IgG4 Concentrations in Patients with Sclerosing Pancreatitis & 2564.4433 &  2001\\
    A &                                                                                                                                                                                 Autoimmune related pancreatitis &          1067.0520 &           2002 \\
    B &                                                                                                                                Close relationship between autoimmune pancreatitis and multifocal fibrosclerosis &            69.1259 &           2003 \\
    C &                                                                                                                                       Elevated IgG4 concentrations in serum of patients with Mikulicz's disease &            70.3783 &           2004 \\
    D &                                                                                                                                       Systemic extrapancreatic lesions associated with autoimmune pancreatitis. &           467.8118 &           2005 \\
    E &                                                                                                                                                       Lymphoplasmacytic sclerosing pancreatitis and cholangitis &            15.6051 &           2002 \\
    F &                                                                                                                              Acute tubulointerstitial nephritis associated with autoimmune-related pancreatitis &           202.6705 &           2004 \\
    G &  IgG4-related sclerosing cholangitis with and without hepatic inflammatory pseudotumor, and sclerosing pancreatitis-associated sclerosing cholangitis: do they belong to a spectrum of sclerosing pancreatitis? &            34.1787 &           2004 \\
    H &                                                                                                                        Long-term prognosis of autoimmune pancreatitis with and without corticosteroid treatment &            16.3812 &           2007 \\
    I &                                                                                                                                                                                         Autoimmune pancreatitis &            35.0497 &           2005 \\
    J &                                                                       Proposal for a new clinical entity, IgG4-positive multiorgan lymphoproliferative syndrome: analysis of 64 cases of IgG4-related disorders &            85.6439 &           2009 \\
    K &                                                                                                                                                                                         Autoimmune pancreatitis &            83.7350 &           2011 \\
    L &                                                                                                                                                                                           IgG4-related disease. &          1068.6365 &           2012 \\
    M &                                                                                                                             Hydronephrosis associated with retroperitoneal fibrosis and sclerosing pancreatitis &            12.7856 &           2002 \\
    N &                                                            Histopathological features of diagnostic and clinical relevance in autoimmune pancreatitis: a study on 53 resection specimens and 9 biopsy specimens &            15.0222 &           2004 \\
    O &                                                                                                                                                   Consensus statement on the pathology of IgG4-related disease. &            11.6600 &           2012 \\
\bottomrule
\end{tabular}
\captionsetup{labelfont=bf}
        \caption{Child article details}
		\label{tab_s3_20} 
\end{table}

\noindent The pioneering work of the topic is `High Serum IgG4 Concentrations in Patients with Sclerosing Pancreatitis'. It was published in 2001, and it already has attracted 1989 citations until 2021. A large number of high knowledge entropy nodes are bred inside the topic, which makes the visible depth of the topic reach 6. \\

\subsection*{S4 MEASURING THE DEVELOPMENT POTENTIAL OF SPECIFIC TOPICS}
\setcounter{table}{0}

%定义编号格式，在数字序号前加字符“A"

\renewcommand{\thetable}{S4-\arabic{table}}

\subsubsection*{Geoscience}
\begin{table}[H]
    \centering
\begin{tabular}{p{1cm}p{12cm}p{2cm}<{\centering}}
\toprule
NO. & Leading article & $\Delta D_{Topic}^{2021}$\\
\midrule
1 & The effect of human mobility and control measures on the COVID-19 epidemic in China & 1.765\\
2 & Mangroves among the most carbon-rich forests in the tropics & 1.746\\
3 & Global, regional, and national comparative risk assessment of 79 behavioural, environmental and occupational, and metabolic risks or clusters of risks in 188 countries, 1990–2013: a systematic analysis for the Global Burden of Disease Study 2013 & 1.445\\
4 & Global land use change, economic globalization, and the looming land scarcity & 1.281\\
5 & Hemispheric and large‐scale land‐surface air temperature variations: An extensive revision and an update to 2010 & 1.225\\
6 & Linear Mixed-Effects Models using `Eigen' and S4 & 1.221\\
7 & The Transiting Exoplanet Survey Satellite & 1.216\\
8 & `Structure-from-Motion' photogrammetry: A low-cost, effective tool for geoscience applications & 1.157\\
9 & Object-based cloud and cloud shadow detection in Landsat imagery & 1.155\\
10 & Bedmap2: improved ice bed, surface and thickness datasets for Antarctica & 1.097\\
\bottomrule
\end{tabular}
\captionsetup{labelfont=bf}
        \caption{Top ten topics in the field of geoscience appearing in the past ten years according to $\Delta D_{Topic }^{2021}$}
		\label{tab_s4_1} 
\end{table}

\noindent In the field of geoscience, the most promising topic is about COVID-19 spatio-temporal data mining. With the COVID-19 pandemic, more and more scientists hope to find a way to fight the epidemic from the spread of the virus. This has led to widespread concern for related research. The list also includes the topics related to mangroves (2), global disease assessment (3) and sustainable development (4).\\

\subsubsection*{Deep learning}
\begin{table}[H]
    \centering
\begin{tabular}{p{1cm}p{12cm}p{2cm}<{\centering}}
\toprule
NO. & Leading article & $\Delta D_{Topic}^{2021}$\\
\midrule
1 & Attention is All you Need & 2.935\\
2 & Prototypical Networks for Few-shot Learning & 2.410\\
3 & Matching networks for one shot learning & 2.229\\
4 & Semi-Supervised Classification with Graph Convolutional Networks & 1.843\\
5 & Understanding deep learning requires rethinking generalization & 1.730\\
6 & A Style-Based Generator Architecture for Generative Adversarial Networks. & 1.723\\
7 & Universal Adversarial Perturbations & 1.705\\
8 & Distillation as a Defense to Adversarial Perturbations Against Deep Neural Networks & 1.567\\
9 & DeepFool: A Simple and Accurate Method to Fool Deep Neural Networks & 1.427\\
10 & Overcoming catastrophic forgetting in neural networks & 1.376\\
\bottomrule
\end{tabular}
\captionsetup{labelfont=bf}
        \caption{Top ten topics in the field of deep learning appearing in the past ten years according to $\Delta D_{Topic }^{2021}$}
		\label{tab_s4_2} 
\end{table}

\noindent In the field of deep learning, the most promising topic is Transformer that proposes the attention mechanism. Transformer originated in the field of natural language processing and is widely used in the field of computer vision. It has gradually developed into a popular deep learning architecture. The research content of the second and third topics (2, 3) are all related to few-shot learning. Currently, deep learning has two important directions. The first is to achieve performance improvement by expanding the parameter scale of the model. The appearance of big models does prove that it is feasible to expand the parameter scale to achieve performance improvement. But it is not the only direction for the development of deep learning. To reduce the dependence on data of deep learning, few-shot learning is another more important research direction. Compared with the improvement of large models' accuracy achieved by hardware accumulation, few-shot learning can bring higher returns. In addition, the list also has pioneering work using graph convolutional networks for semi-supervised classification (4).\\

\subsubsection*{Computer vision}
\begin{table}[H]
    \centering
\begin{tabular}{p{1cm}p{12cm}p{2cm}<{\centering}}
\toprule
NO. & Leading article & $\Delta D_{Topic}^{2021}$\\
\midrule
1 & Rethinking Atrous Convolution for Semantic Image Segmentation & 2.271\\
2 & PointNet++: Deep Hierarchical Feature Learning on Point Sets in a Metric Space & 2.095\\
3 & The SYNTHIA Dataset: A Large Collection of Synthetic Images for Semantic Segmentation of Urban Scenes & 2.057\\
4 & ArcFace: Additive Angular Margin Loss for Deep Face Recognition & 2.044\\
5 & YOLO9000: Better, Faster, Stronger & 2.018\\
6 & You Only Look Once: Unified, Real-Time Object Detection & 2.010\\
7 & Deformable Convolutional Networks & 1.995\\
8 & Image Super-Resolution Using Deep Convolutional Networks & 1.924\\
9 & Dynamic Graph CNN for Learning on Point Clouds & 1.906\\
10 & Image-to-Image Translation with Conditional Adversarial Networkss & 1.892\\
\bottomrule
\end{tabular}
\captionsetup{labelfont=bf}
        \caption{Top ten topics in the field of computer vision appearing in the past ten years according to $\Delta D_{Topic }^{2021}$}
		\label{tab_s4_3} 
\end{table}

\noindent In the field of computer vision, the scientific topics in the first and third place of development potential is related to the semantic segmentation of images. With the urgent need for scene understanding in many practical applications such as current automatic driving and human-computer interaction, inferring corresponding semantic information from images has become a valuable application in the field of computer vision. In addition, there are also groundbreaking topic utilizing neural networks to process point cloud data (2), the creation (6) and improvement (5) of the current important YOLO model in object detection, and the seminal work utilizing deep learning to improve image resolution (8).\\

\subsubsection*{Natural language processing}
\begin{table}[H]
    \centering
\begin{tabular}{p{1cm}p{12cm}p{2cm}<{\centering}}
\toprule
NO. & Leading article & $\Delta D_{Topic}^{2021}$\\
\midrule
1 & A Broad-Coverage Challenge Corpus for Sentence Understanding through Inference & 2.183\\
2 & Get To The Point: Summarization with Pointer-Generator Networks & 1.976\\
3 & Enriching Word Vectors with Subword Information & 1.892\\
4 & SQuAD: 100,000+ Questions for Machine Comprehension of Text & 1.456\\
5 & Listen, attend and spell: A neural network for large vocabulary conversational speech recognition & 1.368\\
6 & Google's Neural Machine Translation System: Bridging the Gap between Human and Machine Translation & 1.319\\
7 & Neural Machine Translation of Rare Words with Subword Units & 1.223\\
8 & Improving Neural Machine Translation Models with Monolingual Data & 1.156\\
9 & BERT: Pre-training of Deep Bidirectional Transformers for Language Understanding & 1.150\\
10 & XLNet: Generalized Autoregressive Pretraining for Language Understanding & 1.091\\
\bottomrule
\end{tabular}
\captionsetup{labelfont=bf}
        \caption{Top ten topics in the field of natural language processing appearing in the past ten years according to $\Delta D_{Topic }^{2021}$}
		\label{tab_s4_4} 
\end{table}

\noindent In the field of natural language processing, the improvement of technology is accompanied by the refresh of various rankings. In our list, the first work is the Multi-Genre Natural Language Inference (MNLI), and the fourth work is the Stanford Question Answering Dataset (SQuAD). This shows that the emergence of various types of datasets has a huge boost to the development of the field. In addition, the popular model Bert in recent years has proved that large models can improve the accuracy of natural language processing tasks and open up a new path for the development of the field, which makes relevant research appear in our list (9, 10).\\

\subsubsection*{Data mining}
\begin{table}[H]
    \centering
\begin{tabular}{p{1cm}p{12cm}p{2cm}<{\centering}}
\toprule
NO. & Leading article & $\Delta D_{Topic}^{2021}$\\
\midrule
1 & Inductive Representation Learning on Large Graphs & 1.560\\
2 & Wide \& Deep Learning for Recommender Systems & 1.598\\
3 & Membership Inference Attacks Against Machine Learning Models & 1.490\\
4 & Deep Neural Networks for YouTube Recommendations & 1.307\\
5 & Neural Collaborative Filtering & 1.259\\
6 & Deep Learning with Differential Privacy & 1.185\\
7 & Graph Attention Networks & 1.185\\
8 & Modeling Relational Data with Graph Convolutional Networks & 1.073\\
9 & XGBoost: A Scalable Tree Boosting System & 1.047\\
10 & Structural Deep Network Embedding & 0.882\\
\bottomrule
\end{tabular}
\captionsetup{labelfont=bf}
        \caption{Top ten topics in the field of data mining appearing in the past ten years according to $\Delta D_{Topic }^{2021}$}
		\label{tab_s4_5} 
\end{table}
\noindent In the field of data mining, the most promising topic is GraphSAGE, which is related to graph neural networks. GraphSAGE solves the problem that the graph convolutional network (GCN) is too slow when representing new nodes, and enables rapid deployment in the production environment to bring practical benefits. In the second place is a recommendation system framework based on deep learning proposed by Google (2). Google has applied this method to its Google Play app recommendation business, and it has also been imitated and applied by many companies. This shows its huge development potential.\\
\end{document}